\documentclass[useAMS,usenatbib]{mn2e}
\pdfoutput=1 
\usepackage{graphicx}
\usepackage{subfigure}
\usepackage{times}
\usepackage{aas_macros}
\usepackage[fleqn]{amsmath}
\newcommand{\WNF}{\text{Panphasia}}
\newcommand{\WNFIAN}{{panphasian}}
\newcommand{\ICgen}{{\sc ic\_2lpt\_gen}}
\newcommand{\MRG}{{\sc mrgk5-93}}
\newcommand{\SUBFIND}{{\sc subfind}}
\newcommand{\PGADGET}{{\sc p-gadget3}}
\newcommand{\GHALO}{{\sc ghalo}}
\newcommand{\msun}{M$_\odot$}
\newcommand{\hmsun}{$h^{-1}$M$_\odot$}
\newcommand{\MDIFF}{{$\Delta R$}}
\newcommand{\GRAFICS}{{\sc Grafic2}}
\newcommand{\DOVE}{{\sc dove}}
\newcommand{\CMBFAST}{{\sc cmbfast}}
\newcommand{\CMB}{{\sc cmb}}
\newcommand{\BAO}{{\sc bao}}
\newcommand{\MXXL}{{\sc mxxl}}
\newcommand{\Millgas}{{\sc mw7}}

\newcommand{\companion}{Jenkins \& Booth (arXiv)}
\def\LCDM{$\Lambda${\sc cdm}}
\def\CDM{{\sc cdm}}
\def\hkpc{$h^{-1}$ {\rm kpc}}

\def\hmpc{$h^{-1}${\rm Mpc}\ }
\def\hgpc{$h^{-1}${\rm Gpc}\ }

\title[A new way of setting Gaussian phases. ]  {A new way
of setting the phases for cosmological multi-scale Gaussian
initial conditions.}

\begin{document}

\author[A. Jenkins]{Adrian Jenkins\thanks{A.R.Jenkins@durham.ac.uk} \\
Institute for Computational Cosmology, Department of Physics, University of Durham, 
South Road, Durham, DH1 3LE, UK}
\maketitle

\begin{abstract} 
 We describe how to define an extremely large discrete realisation of
a Gaussian white noise field that has a hierarchical structure and the
property that the value of any part of the field can be computed quickly.
Tiny subregions of such a field can be used to set the phase
information for Gaussian initial conditions for individual
cosmological simulations of structure formation. This approach has
several attractive features: (i) the hierarchical structure based on
an octree is particular well suited for generating follow up
resimulation or zoom initial conditions; (ii) the phases are defined
for all relevant physical scales in advance so that resimulation
initial conditions are, by construction consistent both with their
parent simulation and with each other; (iii) the field can easily be
made public by releasing a code to compute it -- once public, phase
information can be shared or published by specifying a spatial
location within the realisation.  In this paper we describe the
principles behind creating such realisations. We define an example
called {\sc panphasia}\ and in a companion paper, \companion,
make public a code
to compute it.  With fifty octree levels \WNF\ spans a factor of more
than $10^{15}$ in linear scale -- a range that significantly exceeds
the ratio of the current Hubble radius to the putative \CDM\
free-streaming scale. We show how to modify a code used for making
cosmological and resimulation initial conditions so that it can take
the phase information from \WNF\ and, using this code, we demonstrate
that it is possible to make good quality resimulation initial
conditions. We define a convention for publishing phase information
from \WNF\ and publish the initial phases for several of the Virgo
Consortium's most recent cosmological simulations including the 303
billion particle \MXXL\ simulation.  Finally, for reference, we give
the locations and properties of several dark matter haloes that can be
resimulated within these volumes.

\end{abstract}


\section{Introduction}
 Computer simulations are the main tool for exploring the predictions
of models of cosmological structure formation in the strongly
nonlinear regime.  According to the currently favoured model, \LCDM,
the structure we see today in the Universe is seeded from small
adiabatic Gaussian fluctuations believed to have originated in a very
early inflationary phase in the Universe's history. Nonlinear
structure formation is simulated in \LCDM\ by making initial
conditions at an epoch before there is significant nonlinearity,
followed by integrating the equations of motion forward in time using
for example an N-body code.  The creation of the initial conditions
requires a method to produce realisations of Gaussian random
fields. The techniques to make such realisations have advanced in
tandem with simulation methods over the last three decades.

 Simulation work exploring \CDM\ in the 1980s focused on N-body
modelling of representative volumes of the Universe to study the
large-scale structure in model universes made of collisionless dark
matter.  The Gaussian initial conditions for these simulations were
made by using Fourier methods applied to a single cubic mesh
\citep{Efstathiou85}. These simulations modelled periodic cubic
domains of space sampled with uniform mass particles throughout the
volume.

 The continuing success of the \CDM\ model led in the mid-1990s to the
need to set up more complex simulations to explore the model deeper
into the nonlinear regime including the study of the internal
structure of dark matter haloes.  Because haloes could not be
adequately resolved at reasonable computational cost in cosmological
simulations, new techniques had to be developed to `resimulate' halos
more cheaply. The initial conditions for the resimulations needed both
to faithfully reproduce the target haloes, selected from a
representative cosmological volume, and include extra small scale
power that could not be resolved in the parent simulation. The
computational cost was minimised by coarse sampling the majority of
the simulation volume with massive particles whose function was to
provide the correct tidal environment for the forming halo.  New codes
were developed to make these resimulation initial conditions: to build
multi-mass particle loads and to make multi-scale realisations of
Gaussian fields.

 These first codes, such as the one described in \cite{Navarro_95},
used Fourier methods to generate the multi-scale Gaussian
realisations.  In this example the field is built in a two step
process.  Firstly the Fourier modes present in the original
cosmological simulation are regenerated -- guaranteeing that the
target halo is reproduced. Secondly extra small scale power is
generated. The total displacement field for the particles is the sum
of the contributions from both fields.  The reason for needing two
steps is that it is usually impractical to use a larger Fourier
transform for making the resimulation initial conditions than was used
to generate the initial conditions of the parent simulation.  To get
round this practical limitation the grid for the small scale power is
made physically smaller and placed around just the region that forms
into the halo and its immediate surroundings.  Typically, this second
grid is an order of magnitude smaller than the original grid.  
Over the next decade this method was extended to allow the placing of
further nested grids.  This made it possible to reach very high
numerical resolution in cosmological simulations. Using seven
concentric grids, \cite{Gao05} achieved sub-solar particle mass
resolution for a small patch within a 479~\hmpc\ on a side periodic
volume. 

 While the Fourier method has proved a popular method for making
resimulations for nearly two decades it has, since the late 1990s,
coexisted with an alternative approach for making multi-scale Gaussian
initial conditions suggested by \cite{Salmon96}.  Salmon pointed out
that it is easy to make a discrete realisation of a real-space
multi-scale Gaussian white noise field with the aid of any standard
pseudorandom generator.  This is because the values of a Gaussian
white noise field at different points are independent and so can be
set up sequentially. Once such a white noise field has been generated
it can be transformed by convolving it with an appropriate filter to
produce a Gaussian realisation with any other power spectrum.  This
idea gave birth to what we will call the real-space white noise field
method for making multi-scale Gaussian fields -- it was adopted by
\cite{Pen97} and \cite{Bertschinger01}. Bertschinger released a public
code called \GRAFICS\ based on this idea that can generate multi-scale
Gaussian fields.  This code has been used by others to generate
resimulation initial conditions. A parallel version of \GRAFICS\ has
been developed by \cite{Stadel09} who used it to set up the \GHALO\
series of halo simulations.  Most recently \cite{Hahn_11} have refined
the real-space white noise field method by applying a new algorithm
that uses an adaptive convolution to improve the accuracy of the
numerical convolution by two orders of magnitude when compared to the
Fourier methods deployed in the \GRAFICS\ code.

 To make resimulation initial conditions using a real-space Gaussian
white noise field it is necessary to be able to refine the white noise field
in any region of interest to allow finer spatial scales to be
resolved.  This refined patch of the white noise field must remain
consistent with the unrefined white noise it is replacing to ensure
the same structures are reproduced.  Ideally this process of refining
would operate as if some predefined underlying Gaussian white noise
field were simply being revealed in more and more detail with
increasing levels of refinement rather than simply being invented to
order. This ideal is only partially achieved in the current approaches
to refinement described in the literature. 

 Typically the process of refining starts with a discretised Gaussian
white noise field that is specified on a cubic grid of coarse cells.
Each coarse cell is labelled by a single field value that is
associated with the cell centre.  Each field value is picked at random
from a Gaussian distribution with zero mean and a variance that is
proportional to the inverse of the cell volume.  A refined version of
the field is produced for a grid of finer cells. These fine cells are
generated by splitting the coarse cells in a patch of interest into
smaller equal sized cubic cells.  In \cite{Pen97} and
\cite{Bertschinger01} the Gaussian values over the fine grid are
chosen freely in the same way as the coarse cells, but subject to the
constraint that the sum of the values over the fine cells within each
coarse cell equals the value associated with that coarse cell.  While
elegant, this approach does not guarantee that the corresponding
Fourier modes for the patch as a whole taken in isolation are the same
for the coarse and fine versions.  A more complex approach also using
linear constraints has been developed by \cite{Hahn_11} which forces
the corresponding Fourier modes in the coarse and fine overlapping
region to agree.  While this latter approach guarantees that the
large-scale power on the patch is precisely reproduced in the
refinement, neither method succeeds in defining a truly objective
white noise field -- that is one that is independent of the details of
how the refinements are laid down. This lack of objectivity applies
equally to the Fourier method of making initial conditions as the
precise placement and phases used for the additional grids are
arbitrary.

 The goal of this paper is to develop a practical way to create a
realisation of an objective Gaussian white noise field.  There are two
major advantages for having such a Gaussian white noise field: (i) it
guarantees that sets of initial conditions made for the same region
with different numerical resolutions are as consistent as they can be
-- so that the process of performing successive resimulations becomes
one closer to discovery -- as the structures on all physical scales
are predefined; (ii) if the Gaussian white noise field is made public
then it becomes easy to share or publish the phase information. All
that is needed it to give the precise location of the phase
information within the realisation as a whole. The routine publishing
of the phase information has the potential to enrich the literature by
making it easier for others to check, reproduce or build upon
published simulation work.

  The reason Gaussian white noise fields are convenient to work with
numerically is due to one particular property. This is the fact that
the expansion coefficients of a Gaussian random white noise field with
respect to any orthogonal basis function expansion are necessarily
independent Gaussian variables. This property means that a realisation
of Gaussian white noise fields can be conveniently created by using a
pseudorandom number generator to set the values of the expansion
coefficients.

The fact that this property is true for any orthogonal basis function
set suggests that it might be possible to find a set of orthogonal
basis functions that are particularly well suited for making
cosmological initial conditions. Neither Fourier modes nor sets of
independent values arranged  on a real space grid,  are
obviously optimal for the task of generating cosmological multi-scale
Gaussian fields.

 The ability to resimulate any region of a simulation to any desired
numerical resolution requires that it must be possible to successively
refine the Gaussian white noise field at an arbitrary location.  This
requirement suggests looking for an orthogonal basis set with a
hierarchical structure.  An octree, which is the set of cells formed
by dividing a cube into eight sub-cubes and continuing this operation
recursively on the sub-cubes, would seem the simplest and most
convenient geometrical structure to adopt.  Using this structure we
can define `orthogonal octree basis functions' to be functions that
are localised to particular octree cells being zero everywhere else.
Clearly defined this way octree basis functions in octree cells that do not
overlap are trivially orthogonal. For octree cells that do overlap the
requirement of orthogonality is non-trivial, and limits the possible
functional forms for the octree basis functions as we show later. The
high symmetry and self-similar nature of an octree means in practice
that a relatively small set of functional forms are needed to describe
the infinite set of octree basis functions needed to populate an
octree to unlimited depth.

 We can define the phase information for a periodic cosmological
simulation in the following way.  We identify the cubic volume of the
simulation with a corresponding cubic sub-volume within the octree.
This region in the octree can be made of a single octree cell, or a
group of cells.  We can then use the values of the Gaussian white
noise field in this chosen sub-volume of the octree to define the
phases for the cosmological simulation to any desired resolution.
Because a Gaussian white noise field within a region is completely
independent of the field outside of that region, we can effectively
cut out conveniently sized cubic blocks from a much large Gaussian white
noise field and use these independent blocks to define the phases for
particular cosmological volumes.

The octree functions form a discrete four dimensional space -- three
of these dimensions span physical space in the form of a cubic grid
consisting of eight to some integer power cells, while the fourth
dimension spans the allowed side lengths of the octree cells which are
the side length of the root cell divided by two to some integer power.
A realisation of a Gaussian white noise field can be made using these
octree basis functions by first establishing a 1-d to 4-d mapping
between a pseudorandom Gaussian number sequence and the space of
octree functions.  Each pseudorandom number is taken as the expansion
coefficient of the white noise field for a particular octree basis
function.  Similar mapping strategies, although more commonly from 1-d
to 3-d, are used in all methods used to make Gaussian cosmological
initial conditions.

 To exploit the full potential of this 1-d to 4-d mapping so it is
possible to refine the white noise field at any location to any depth,
it must be possible to access all of the relevant expansion
coefficients at reasonable computational cost.  This requires choosing
a pseudorandom number generator that allows large jumps through the
linear pseudorandom sequence to be made cheaply.  Fortunately there
are classes of pseudorandom number generators with this property, and
amongst these there are well tested generators in common use.

 Given that it is possible to access any expansion coefficient
relatively easily, it becomes possible by assigning the entire period
of the pseudorandom generator to the octree to create a realisation of
a Gaussian white noise field with a truly enormous dynamic range. The
typical periods of generators commonly in use are so large that the
resulting white noise field is far larger than needed for any one
simulation, or indeed for all simulations that have ever been run (at
least on this planet!). For most generators the period is easily big
enough to define a white noise field that can resolve scales below the
putative \CDM\ free streaming scale \citep{Hofmann_01} everywhere
within a volume that greatly exceeds the current Hubble volume.

  The rest of this paper is a detailed elaboration of the ideas
outlined in this introduction leading to the construction of a
particular realisation, called \WNF, which is designed for the purpose
of making accurate cosmological and resimulation initial conditions.
We make this field public in a companion \companion. The outline of
the rest of the paper is as follows: in Section~\ref{MATH-INTRO} we
will give the mathematical background to the properties of Gaussian
fields needed later.  In Section~\ref{OCT-GEN} we give a general
description of how to construct the octree orthogonal basis functions
and outline their properties and choose the most suitable set for
making simulation initial conditions. We save the nitty-gritty and
more tedious details of the practical implementation to
Appendix~\ref{OCT-DET}.  In Section~\ref{FIND-GEN} we introduce the
pseudorandom number generator and its properties, but leave the
details of the precise mapping of the sequence to the octree to
Appendix~\ref{MAP-OCT}.  In Section~\ref{ADD-ICGEN} we show how to add
\WNF\ to the \ICgen\ initial conditions code first described in
\cite{Jenkins2010}. In Section~\ref{SIM-TESTS} we generate and test
cosmological and resimulation initial conditions and show that it is
possible to obtain good results using \WNF.  In
Section~\ref{PUB-PHASE} we define a convention for publishing phases
and give the phases for several of the most recent Virgo Consortium
volumes together with the locations of a few haloes within these
volumes.  In Section~\ref{CODE-DES} we give an overview of the code to
compute \WNF.  Section~\ref{SUM-PAP} is the summary. Finally in
Appendix~\ref{PANPHASIA-DEF} we give the formal definition of \WNF.

\section{Mathematical background\label{MATH-INTRO}}
\subsection{Orthogonal basis function expansions of Gaussian white noise 
fields\label{Gauss_fields}} In the \LCDM\ model the primordial
 density fluctuations are a homogeneous and isotropic Gaussian
 field. Taking the spatial curvature to be negligible, a \LCDM\
 universe can be modelled as a finite cube of side length $L$, with
 periodic boundary conditions. For such a cube we can describe the
 density fluctuations in terms of the matter overdensity,
 $\delta({\bmath x})$, where $\bmath x$ is the position, as a sum over
 Fourier modes:
\begin{equation}
    \delta({\bmath x}) = \sum_{\bf k} \delta({\bmath k})
\exp\left[i{\bmath k.\bmath x}\right].
\label{Gaussian_field}
\end{equation}
The periodic boundary conditions require the wave vector, $\bmath
k$, to take discrete values with Cartesian components
$(k_x,k_y,k_z) = (2\pi/L)(l_x,l_y,l_z),$ where $l_x,l_y,l_z$ are
integers.

  By definition a Gaussian field is one where the amplitudes of the
Fourier modes, $\delta({\bmath k})$, are independent, with the real and
imaginary parts of each mode drawn from the same Gaussian
distribution.  The statistical properties of a Gaussian random field
are completely determined by its power spectrum, which is defined by:
\begin{equation}
P({\bmath k}) = \langle|\delta({\bmath k})|^2\rangle/L^3,
\end{equation}
where the angled brackets signify an ensemble average.
 For a real Gaussian field with zero mean overdensity, the Fourier
mode amplitudes obey the constraints: $\delta({\bmath k} = (0,0,0)) =
0$, and $\delta({\bmath k}) = \delta^*(-{\bmath k})$, where
$\delta^*$ is the complex conjugate of $\delta$.

By definition a Gaussian white noise field has a constant power
spectrum. For convenience we will take $\langle|\delta({\bmath
k})|^2\rangle=1$ for all Gaussian white noise fields in this paper.
There are two properties of Gaussian white noise fields that are
particularly relevant for this paper.

 Firstly, a white noise field has power at all wavenumbers which means
that it can always be transformed into another Gaussian field with any
desired power spectrum by convolving it with a suitable kernel
function.  As is well known, the operation of a spatial convolution
corresponds in Fourier space to a simple scaling of the amplitude of
each of the Fourier modes.  By choosing a scaling factor with a
modulus of $\sqrt{P(\bmath k)}$ the power spectrum of the convolved
white noise field becomes $P(\bmath k)$. The scaling can include an
arbitrary phase factor, but as the goal of this paper is to use a
white noise field to define the phase, we need to insist that the
kernel function must be real and non-negative in Fourier space.  With
this choice the phase information is purely contained in the white
noise field itself.

 The second property, which we derive here, is the fact that the basis
function coefficients of any orthogonal basis function expansion of a
Gaussian white noise field are necessarily independent Gaussian
variables.  Our definition of a Gaussian field asserted that the
expansion coefficients are independent Gaussians for a plane wave
expansion, and the plane waves are an example of an orthogonal basis
function set. We now show that given this is true for a
plane wave expansion of a Gaussian white noise field, it is also true
for all other sets of orthogonal basis functions.

To show this we first consider a vector $\bmath V$ with $N$
components, $v_i,i=1,N$, which are Gaussian independent random
variables with $\langle v_i \rangle = 0$, and $\langle v_iv_j\rangle =
\delta_{ij}.$ When we say the variables are independent we mean that
the joint probability distribution of the $N$ variables is:
\begin{equation}
  {\rm Prob}(v_1,v_2,\cdots,v_n) = \frac{1}{(2\pi)^{N/2}}\exp\left[-\frac{1}{2}\bmath V^T\bmath V\right], 
\label{vjointprob}
\end{equation}
which is just the product of the $N$ individual Gaussian probability distributions. 

Consider now an alternative vector $\bmath W$, with $N$ components
$w_i,i=1,N$ which are linearly related to the $v_i$ by 

\begin{equation}\bmath W =
\mathbfss{R}\bmath V, 
\label{orthog_trans}
\end{equation}
where the matrix $\mathbfss{R}$ is any orthonormal matrix, so that
$\mathbfss{R}^T\mathbfss{R} = \mathbfss{I}$.

 As the magnitude of the Jacobian for a linear transformation between
the $v_i$ and $w_i$ variables is just the magnitude of the determinant
of $\bmath R$, which is unity for an orthonormal matrix, we can simply
transform eqn~(\ref{orthog_trans}) to give the joint probability
distribution of the $w_i$:
\begin{equation}
  {\rm Prob}(w_1,w_2,\cdots,w_n) = \frac{1}{(2\pi)^{N/2}}
\exp\left[-\frac{1}{2}\bmath W^T\bmath W\right],
\label{wjointprob}
\end{equation}
 which shows that the $w_i$ are independent Gaussian variables too.

To apply this result to Gaussian white noise fields themselves we need
to go from using finite vectors to using infinite vectors to represent
functions in Hilbert space, in the manner made familiar by quantum
mechanics.

 Let the $f_i({\bmath x})$ be an infinite set of real
functions defined over the periodic volume which  
obey these orthogonality and normalisation relations:
\begin{equation}
 \int f_i({\bmath x})  f_j({\bmath x}) {\rm d^3}{\bmath x} = \delta_{ij},
\label{forthog}\end{equation}
 where the indices, $i$, $j$ label the functions and the integral is over
the volume.  
Using Parseval's relation the corresponding
normalisation/orthogonality relations in Fourier space are:
\begin{equation}
 \sum_{\bmath k} \tilde{f}_i({\bmath k})  \tilde{f}_j({\bmath k})  = \delta_{ij},
\label{Fourier_orthog_rel}
\end{equation}
 where the function $\tilde{f}_i$ is the Fourier transform of $f_i$.
  If we express a Gaussian white noise field as an expansion in the
functions $f_i$, with expansion coefficients $C_i$, and assume that
these functions form a complete set then we can express the overdensity
field as a sum over these basis functions:
\begin{equation}
    \delta({\bmath x}) = \sum_j C_jf_j({\bmath x}), 
\label{func_exp}\end{equation} 
where the sum is over the whole set of functions.

 Formally this sum is ill defined for a white noise field, 
but nonetheless this
expression can be employed inside of integrals in a well defined way.
An expression for the $C_i$ can be obtained by multiplying both sides by
$f_i$ and integrating over all space, applying Parseval's
relation to the l.h.s.  and eqn~\ref{forthog} to the r.h.s to give:
\begin{equation}
    C_i = \sum_{\bmath k} \tilde{f}_i({\bmath k}) \delta({\bmath k}), 
\label{new_func_exp}\end{equation}
where the sum is over all wavenumbers. We can recognise this equation
as being an orthonormal transformation, analogous to
eqn~\ref{orthog_trans} between the plane wave set of expansion
coefficients, $\delta({\bmath k})$, and the $C_i$ expansion
coefficients of the $f_i$ orthogonal basis function set.  Thus, the
$C_i$ must be independent Gaussian variables.

 Having shown this, we can define a Gaussian white noise field in terms
of the $C_i$ coefficients. By using a pseudorandom generator to assign
values to the $C_i$ we can create a realisation of a Gaussian white
noise field in terms of the $f_i$ functions.

To date only two sets of orthogonal basis functions have been used for
making cosmological simulation initial conditions. These are the plane
waves and real-space grids of delta functions.

 Plane waves are attractive because it is possible to refine a
Gaussian white noise field simply by adding higher wavenumber modes to
the modes already present.  Their disadvantage is that this approach
is very expensive computationally. This is because the plane waves are
not localised in real space.  This means to refine a white noise field
in one region it is necessary to provide the information to refine the
field everywhere to the same degree.  This high cost has been avoided
by for example \cite{Navarro_95} by placing a smaller Fourier grid
around the region of interest so that the newly added modes only
contribute to a small part of the simulation volume.  This approach is
necessarily approximate as the added modes are not truly orthogonal to
the original plane waves.

 By contrast,  arrays of delta-functions are perfectly localised, which
makes them very efficient from a computational point of view. However,
their disadvantage is that they are not able to resolve scales smaller
than the distance between adjacent delta-functions and cannot simply
be added to to generate a finer version of the field.  The approach in
the literature \citep{Pen97, Bertschinger01, Hahn_11} to get round
this limitation has been to simply replace one set of delta-functions
with an `equivalent' finer set. This equivalence is achieved by using
a set of linear constraints to force the phase information to be as
similar as possible between the original and its replacement. This
approach is also approximate.

  However, it is possible to find orthogonal basis functions that have
both a strong degree of locality like the real-space delta functions,
and that enable a Gaussian white field to be refined just by adding
more components as with the plane waves. In this paper we develop
a set of orthogonal octree basis functions that have these properties.
The octree basis functions are spatially extended, but each is
localised to a single cell of the octree.   Using them a Gaussian
white noise field can be refined at relatively low computational
cost by just adding new basis functions from deeper in the octree
to the existing field in the region of interest only.
 
Before discussing the nature of the basis functions we first define
some terminology to describe the octree structure we need.

\subsection{Notation to describe an Octree\label{octree}}
  We identify the root cell of the octree with the whole periodic cubic
spatial domain of length $L$. We define a set of Cartesian coordinates,
$(x_1,x_2,x_3)$, 
aligned with the three orthogonal edges of the root cell and place the corner at
$(0,0,0)$. The coordinates have an allowed range:   $0\le x_i <L$, $i=1,2,3$.
We define the root cell to be at level 0 of the octree.

  At level $l$ of the octree there are $8^l$ cubes each of side-length:
\begin{equation}
\Delta_l =\frac{L}{2^l}.
\label{deltal}\end{equation}
 We will label these cells using
integer Cartesian coordinates, $j_1,j_2,j_3$ where $0\le j_i < 2^l$,
$i=1,2,3$.  The centre of a cell $(j_1,j_2,j_3)$  
at level $l$, $\bmath x_c$ has coordinates: 
\begin{equation}
  {\bmath x_c}(l,j_1,j_2,j_3) = ( j_1+1/2,j_2+1/2,j_3+1/2)\Delta_l.
\label{cell_centre}\end{equation}
   Each octree cell has eight child cells, and we will call this cell
the parent cell with respect to any of its child cells.

\section{Building the orthogonal octree basis functions.\label{OCT-GEN}}
 The primary objective of this paper is to use a new orthogonal basis function set
based on an octree structure to construct a realisation of a Gaussian
white noise field.  By definition these octree basis functions are
localised to particular octree cells and mutually orthogonal. We
require that the infinite set of octree basis functions can be
described by a finite set of functional forms that are common to all
levels of the tree, save for a normalisation constant that is allowed
to depend on the level of a cell in the octree.

 It proves convenient numerically to define the octree basis functions
themselves in terms of a smaller set of more primitive building block
functions.  The octree basis functions can all be built from different
combinations of these building blocks.  These building block functions
are similarly localised to particular octree cells and, as described
later, are each separable into the product of three one dimensional
functions of the Cartesian coordinates.  An approximation to a
Gaussian white noise field can be constructed using these blocks by
placing one or more of these functions into each octree cell at level
$l$ of the octree so that they collectively tessellate the entire
volume.  Each function is given an individual weight, which is
determined by the Gaussian white noise field.  For convenience, the
functional forms of these building blocks are chosen so that they form
an orthogonal basis set when occupying the octree cells at a single
level of the octree. We can use these blocks to approximate a Gaussian
white noise field by making a basis function expansion of the field
with these blocks. The expansion coefficients of these blocks will be
independent Gaussian variables as the blocks are orthogonal.

This representation is an approximation to a white noise field because
no finite set of building block functions can form a complete basis
set. For cosmological initial conditions we require that the density
fluctuations be accurate from large scales down to some minimum
lengthscale determined by numerical reasons such as by the particle
Nyquist frequency or the gravitational softening length.  For this
basis function expansion of a Gaussian white noise field to be a
useful approximation for our purposes, we require that the power
spectrum tends to unity in the limit $k\rightarrow0$ so that the power
on large scales is accurately represented.  If this condition is met
then it becomes possible to generate an accurate approximation of a
Gaussian white noise field for all wavevectors with a modulus below
some given value provided the building block functions are placed at a
sufficient depth in the octree. It is not hard to see that a building
block function that is just a constant within an octree cell and zero
everywhere else is one possibility.  As we will see later we can add
additional building block functions to improve the rate of convergence
to a white noise spectrum in the limit of $k\rightarrow0$.

  Another practical requirement on the possible set of building block
functions is that is must be possible to construct each of these
functions when placed in an octree cell at level $l$, out of a
superposition of the same set of building blocks placed at level $l+1$
in the eight child cells.  If this is true and we assume that the
octree functions at level $l$ can be built from the building block
functions placed at some deeper level, then it follows that all of the
octree basis functions from level $l-1$ to the root cell can similarly
be exactly represented by these building block functions at this one level.  In
other words, it is possible to project the four dimensional space of
octree basis functions, above any given depth, onto a three
dimensional grid and represent it exactly using the building block
functions placed in these grid cells with appropriate weightings.

 In fact the octree basis functions are to a large extent implicitly defined
by the choice of these building block functions. To see this imagine
taking a given realisation of a Gaussian white noise field and
approximating it in two different ways as a basis function expansion
in a given set of building blocks at either level $l$, or at level
$l+1$. The expansion at level $l+1$ will contain all the information
present in level $l$ expansion, but not vice versa.  The additional
information about the white noise field at level $l+1$ is just that
introduced by the addition of a single layer of octree basis
functions.  It follows from this that the octree functions, which
occupy whole octree cells at level $l$, must each be built of eight
blocks with each child cell containing some superposition of the
building block functions and that they are orthogonal to the level $l$
building blocks. As the expansion at level $l+1$ has eight times the
number of expansion coefficients as at level $l$, it follows that the
number of independent octree functions must be seven times the number
of building block functions.  By a similar argument we can deduce that
the ensemble average power spectrum of a layer of octree basis
functions placed at level $l$ of the octree is simply given by the
difference between the ensemble average power spectrum of a white
noise field expanded using the building blocks at levels $l$ and
$l+1$.

 While it may not be immediately obvious, it is not hard to see that
it is possible to create sets of building blocks with the required
properties starting from Legendre polynomials.

\subsection{Using Legendre functions to build the octree basis 
functions\label{building_blocks}}
 The Legendre polynomials, by definition, are localised in a finite interval, and 
 obey the orthogonality relations:
\begin{equation}\label{lpor}
 \int_{-1}^1 P_l(x)P_m(x){\rm d}x = \frac{2}{2l+1}\delta_{lm}.
\end{equation}
 The lowest order Legendre polynomials are: $P_0(x) = 1;  P_1(x) = x; 
P_2(x) = (3x^2-1)/2$. 
 
 We can define three dimensional `Legendre block' functions as
products of Legendre polynomials in all three Cartesian coordinates
within a unit length cubic cell and zero outside as follows:
\begin{equation}\label{legendre_block}
  p_{j_1j_2j_3}({\bf x}) = 
   \begin{cases}
       \displaystyle \strut   \displaystyle \strut\prod_{i=1}^3  
          (2j_i+1)^{1/2}P_{j_i}(2x_i) & \text{if  $-\frac{1}{2}\le x_i < \frac{1}{2}$;} \\
                 0                                       & \text{otherwise;} \\
   \end{cases}
\end{equation}
 The Legendre blocks, when placed within octree cells at a given 
level in an octree, are orthogonal and
obey the following normalisations and orthogonality relations when
integrated over all space:
\begin{equation}
    \int p_{i_1i_2i_3}({\bf x}) p_{j_1j_2j_3}({\bmath x}){\rm d}^3{\bmath x} = \delta_{i_1j_1}
      \delta_{i_2j_2}\delta_{i_3j_3}.
\end{equation}

 All of the Legendre blocks meet the requirement for the building
blocks to be localised and orthogonal. The requirement that each block
can be built of a superposition of blocks deeper within the octree
does place some constraints on the suitable sets of blocks. We will
refer to a set of blocks with this property as being self-representing.

 It is not hard to see that it is possible to define sets of blocks
that are self-representing by taking all possible blocks built from
combinations Legendre polynomials up to some given order.  We will
call the S$_1$ set the set that made from $P_0$ alone, S$_8$ that made
from the eight combinations of $P_0$ and $P_1$ polynomials and
S$_{27}$ that made from the twenty-seven combinations of $P_0$, $P_1$
and $P_2$ polynomials.  We will refer to the S$_1$, S$_8$ and
occasionally the S$_{27}$ sets of Legendre blocks throughout the rest
of the paper.  We will use these labels also to refer to the octree
basis function themselves, so when we refer to for example the S$_8$ octree
basis functions, we mean those that are built from the S$_8$ set of
Legendre block functions.

 Having defined some potential sets of Legendre blocks for building a
Gaussian white noise field we need some way to judge their relative
merits. While all Legendre blocks contribute equally to the total
variance of the white noise field (in an ensemble averaged sense),
they differ in their relative contributions to the power spectrum as a
function of wavenumber. To examine this further we need to look at the
Fourier representations of the Legendre blocks.

 The Fourier transform of the Legendre polynomials in
the $[-1,1]$ interval are the Spherical Bessel functions.  The lowest
order Spherical Bessel functions are $j_0(k) = \sin(k)/k$, $j_1(k) =
(\sin(k)-k\cos(k))/k^2$. As $k\rightarrow0$ so $j_n(k)\rightarrow k^n/(2n+1)!!$. 
We define the Fourier transform of the
Legendre blocks as follows:
\begin{equation}
 \int_{L^3} p_{j_1j_2j_3}({\bmath x})\exp[i{\bmath k}.{\bmath x}]{\rm d}^3{\bmath x} 
= i^n j_{j_1j_2j_3}({\bmath k}), 
\label{sph_bess_fn_def}
\end{equation}
  where the functions, $j_{j_1j_2j_3}$ are related to the Spherical Bessel functions:
\begin{equation}
  j_{j_1j_2j_3}({\bf k}) = \prod_{i=1,3}  (2j_i+1)^{1/2} j_{j_i}(\frac{k_i}{2}). 
\end{equation}
 The Spherical Bessel functions obey an identity for all $k$:
\begin{equation}
  \sum_{l=0}^\infty (2l+1)j_l^2(k) = 1.
\end{equation} 
 Similarly the functions  $j_{j_1j_2j_3}$ obey:
\begin{equation}
   \sum_{j_1 = 0}^\infty \sum_{j_2 = 0}^\infty  \sum_{j_3 = 0}^\infty j^2_{j_1j_2j_3}({\bmath k}) = 1.
\label{sph_bessel_identity}\end{equation}
We will need this identity later to establish the completeness of the octree
basis functions.

It is possible to have two self-representing sets of Legendre blocks,
S$_M$ and S$_N$ with $M<N$, where the former set of blocks is a subset
of the latter.  For example S$_8$ is a subset of S$_{27}$.  In such
cases given a realisation of a Gaussian white noise field that has
been constructed using the S$_N$ octree basis functions it is possible
to obtain an equivalent representation of exactly the same field but
built from the S$_M$ octree basis functions by simply ignoring those
Legendre blocks that are not part of S$_M$. While this might seem
paradoxical, this equivalence is only true for expansions made with
complete sets of octree basis functions which are infinite in number.
For expansions made with a finite set of octree basis functions some
choices are better than others as judged for the purposes of making
initial conditions.  We will show this later by comparing simulations
of a particular dark matter halo at redshift zero run from initial
conditions made using either S$_1$ or S$_8$ octree basis functions.

 The reverse procedure of starting with a field based on the set
S$_M$, and wanting to create an equivalent field but using the
superset S$_N$ is non-trivial. The coefficients of the blocks that are
part of S$_N$ but not S$_M$ are implicitly determined by an infinite
number of coefficients belonging to the S$_M$ blocks at deeper levels
of the tree.  For this reason it is better to be somewhat conservative
in the initial choice of sets of Legendre block functions and to try
and take as large a set as might possibly be needed.  Taking too large
a set however risks making the generation of the field needlessly
slow.  For this paper we will evaluate just the S$_1$ and S$_8$ octree
basis functions.  As will be shown later the former does not perform
very well and the latter performs well enough that there is no
compelling reason to look at more elaborate choices. 

\begin{figure}
\resizebox{\hsize}{!}{
\includegraphics{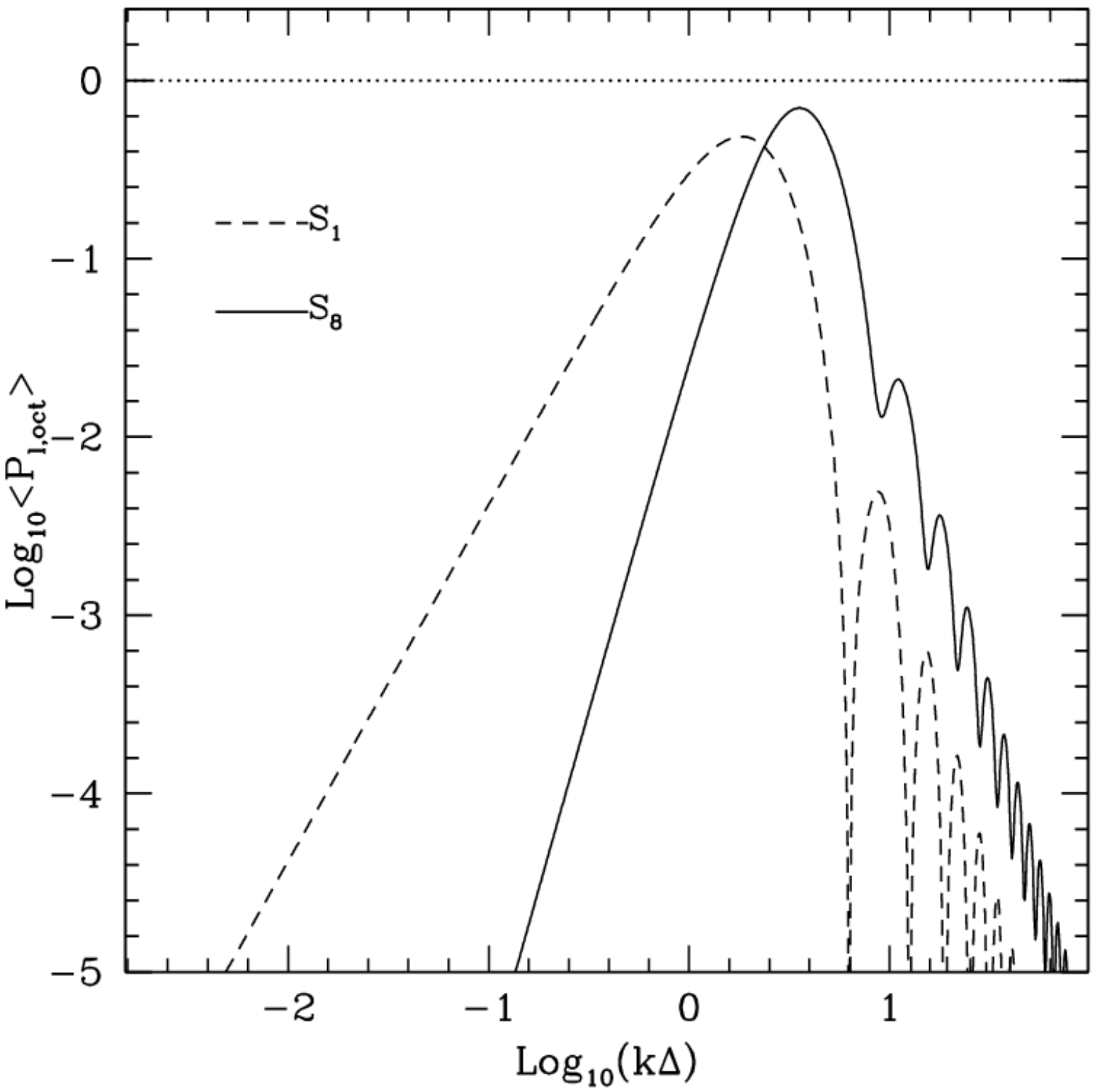}}
\caption{ The ensemble average power spectrum of the octree
basis functions, given by eqn~(\ref{octree_power}),  at a single level of the octree for two
different choices of Legendre blocks. The power spectrum
given is the average over a cubic shell centred on the 
origin and labelled by the maximum Cartesian coordinate.
The horizontal dotted line is the amplitude of the white
noise field.  Both S$_1$ and S$_8$ have a logarithmic slope of -4 at
large values of $k\Delta$. The logarithmic slopes at small values
are 2 and 4 respectively.
}
\label{octree_powspec}
\end{figure}

\subsection{Properties of the octree basis functions.}
 We can write the ensemble power spectrum of a basis function
expansion of a Gaussian white noise field using the set S$_N$ of Legendre
blocks at level $l$ of the octree as:
\begin{equation}
\langle P^N_l({\bf k})\rangle = \sum\limits_{{\rm S}_N} 
j^2_{j_1j_2k_3}({\bf k}\Delta_l),
\label{level_power}\end{equation}
where  $\Delta_l$ is the size of the octree cells, defined in eqn~\ref{deltal}.
 
 The $j_{000}$ function, present in all sets, tends to unity as
$k\rightarrow0$, while the sum over the whole set of Legendre
block functions, given by eqn~\ref{sph_bessel_identity}, is unity for
all wavenumbers. We can therefore see that the power spectrum
approaches unity from below as ${\bf k}\Delta_l\rightarrow0$.  Similarly the
power spectrum tends to unity from below for any given ${\bf k}$ as
$\Delta_l\rightarrow0$.  This demonstrates that the set of octree
functions is complete: the Fourier mode set is complete and  
the amplitude of any Fourier mode can be reproduced by the octree
basis functions in the limit of infinite tree depth.

  For practical purposes the tree depth is finite, and there are
significant differences between how well different sets of octree
basis functions approximate the large-scale modes of a white noise
field.  As mentioned at the beginning of this section, we can
determine the ensemble power spectrum of a single level of octree
basis functions by taking the difference between the ensemble average
power spectrum of the building blocks placed at two adjacent levels of
the octree:
\begin{equation}
\langle P^N_{l-1{\rm ,oct}}({\bf k})\rangle = \sum\limits_{{\rm S}_N} 
\left(j_{j_1j_2j_3}^2({\bf k}\Delta_l)
                          - j_{j_1j_2j_3}^2(2{\bf k}\Delta_l)\right).
\label{octree_power}
\end{equation}
 Note we have implicitly associated a level $l-1$ for the octree basis
functions that are made from eight level $l$ Legendre block
functions. 
 
 Figure~\ref{octree_powspec} shows the ensemble average octree power
spectrum for a single octree layer for the S$_1$ and S$_8$ sets of
Legendre blocks. The power spectrum shown is an average over cubic
shells in $k$-space.  The power in S$_8$ is more sharply peaked with the
logarithmic slope at low $k\Delta$ of 4, compared to 2 for S$_1$.  At
high $k\Delta$ the logarithmic slopes of both are -4, but as the S$_8$
set contains eight times as many basis functions as S$_1$, it has
eight times more power and therefore has a significantly higher
amplitude at higher $k$.  The three dimensional power spectrum of the
octree functions has cubic symmetry and is therefore anisotropic. We
will leave the practical issue of how to restore isotropy on all
scales to later in the paper when we describe how to make initial
conditions.

 The deviation of the power spectrum from unity at low $k$ for the
S$_8$ set scales as $k^4$ which is significantly better than the $k^2$
scaling of the S$_1$ set.  This makes the S$_8$ set significantly
better at approximating the large-scale power. By this measure the
S$_{27}$ set would be even better with its $k^6$ scaling, but there
are disadvantages in using large block sets.

 Using a larger set of Legendre blocks incurs a greater computational
 expense when evaluating the field. This cost is made up of two
 components: the extra pseudorandom numbers that need to be computed,
 which scales linearly with the number of Legendre blocks, and the
 time to compute the relevant Legendre coefficients from the octree
 basis functions which involves linear algebra with a matrix whose
 size scales as the square of the number of Legendre blocks.  For the
 S$_8$ set these two elements take similar amounts of cpu time. We
 would expect that S$_{27}$ would be about a factor of roughly ten more
 expensive than S$_8$ to evaluate per octree cell. For
 practical reasons, discussed later, it would also be necessary to
 evaluate the S$_{27}$ field more times to avoid the code being any
 more memory intensive and that could make it 30 times more expensive.
 This view is informed by the performance of the code which we make
 public in \companion. If a significantly faster code could be
 developed to compute the field then the practical argument against
 using S$_{27}$ set would be weakened.

 In the next subsection we explicitly define sets of S$_1$ and
S$_8$ octree basis functions.

\subsection{Functional forms for the S$_1$ and S$_8$ octree functions.}

  While the octree basis functions are three dimensional functions,
they can be factorised into products of three one dimensional
functions of each of the Cartesian coordinates. For S$_1$ these
one dimensional functions are built from the $P_0$ Legendre polynomial,
which is just a constant, while for S$_8$ the $P_0$ and $P_1$
Legendre polynomials are required.
For S$_1$ we define two one dimensional functions:
\begin{equation}
  D_0(u) = \begin{cases}  \phantom{-}1 &  \text{if $ -1\leq u < 1$;} \\
                         \phantom{-}0 &  \text{otherwise,}
         \end{cases}
\end{equation}
\begin{equation}
  D_1(u) = \begin{cases}   \phantom{-}1 &  \text{if $ 0\leq u < 1$;} \\
                                   -1 &  \text{if $ -1 < u < 0$;}  \\
                         \phantom{-}0 &  \text{otherwise.}
         \end{cases}
\end{equation} 
Using these functions we can generate eight three dimensional
functions, $F_{ijk}$, occupying an octree cell at level $l$ as
follows:
\begin{equation}
  F_{ijk}^l({\bmath x}) =  \frac{1}{\Delta^{3/2}_l} D_i\left(\frac{2x_1}{\Delta_l}\right)
   D_j\left(\frac{2x_2}{\Delta_l}\right) D_k\left(\frac{2x_3}{\Delta_l}\right),
\label{sone-funcs}\end{equation}
 where $i$, $j$, $k$ are the integers either zero or one, $x_1$,
$x_2$, $x_3$ are the Cartesian components of ${\bmath x}$ and it is
assumed the origin is the centre of the octree cell at level $l$.  We
can consider all of these functions of being built from eight 
$p_{000}$ Legendre blocks placed, with appropriate weights,
in the child cells at level $l+1$.

  The function $F^l_{000}$ is just a constant and corresponds to a
$p_{000}$ Legendre block at level $l$. The seven other functions are
the octree basis functions themselves. Note that each octree function
has at least one discontinuity in value.  Given that the functions
$D_0$ and $D_1$ are respectively symmetric and antisymmetric about the
origin it follows that all eight functions are mutually orthogonal:
\begin{equation}
 \int F^l_{ijk}({\bmath x}) F^l_{lmn}({\bmath x}) {\rm d}^3{\bmath x} = \delta_{il}
   \delta_{jm}\delta_{kn},
\end{equation}
 when integrated over the volume of the cell at level $l$. Clearly all
seven octree basis functions within a given octree cell are mutually
orthogonal. It is easy to see, given that these seven functions are
orthogonal to the $p_{000}$ Legendre block at the same level, that all
octree basis functions, no matter what octree cells they occupy, are
mutually orthogonal. The functional forms of the octree basis
functions given in the equation above are particularly simple, but
they are not unique.  Alternative functions can be generated by using
any $7\times7$ orthonormal matrix, as in eqn~\ref{orthog_trans} to
produce a new set.

 It is not hard to see that if a given realisation of a Gaussian white
noise field is expanded using the $p_{000}$ Legendre blocks at both level
$l$ and at $l+1$, then the expansion coefficient of each block at
level $l$ is just the sum of the corresponding eight coefficients of
its child cells.  This relationship between the parent and child
coefficients is identical to that used by both \cite{Pen97} and
\cite{Bertschinger01} for refining a real space Gaussian white noise
field.

  From a coding point of view it is tempting to add \WNF\ to \GRAFICS\
using only the information in S$_1$ block coefficients as this would
be quick and easy to do.  However, as we will see in the tests shown
later in this paper using the S$_1$ block alone does a poor job in
reconstructing the \WNF\ phase information, so we cannot recommend
this approach.

  We now define a set of S$_8$ octree basis functions.  These are
built from the eight Legendre block functions which are products of
the zeroth and first Legendre polynomials. We can do this in an
analogous fashion to eqn~\ref{sone-funcs} by first defining a set of
four one dimensional functions, $E_i$,  that are the S$_8$ analogs of the
two $D_0$ and
$D_1$ functions used to define the S$_1$ octree basis functions.
These four functions are built from combinations of the $P_0$ and
$P_1$ Legendre polynomials. Similarly for S$_{27}$ we would need
six functions built from $P_0$, $P_1$ and $P_2$.  

\begin{figure}
\resizebox{\hsize}{!}{
\includegraphics{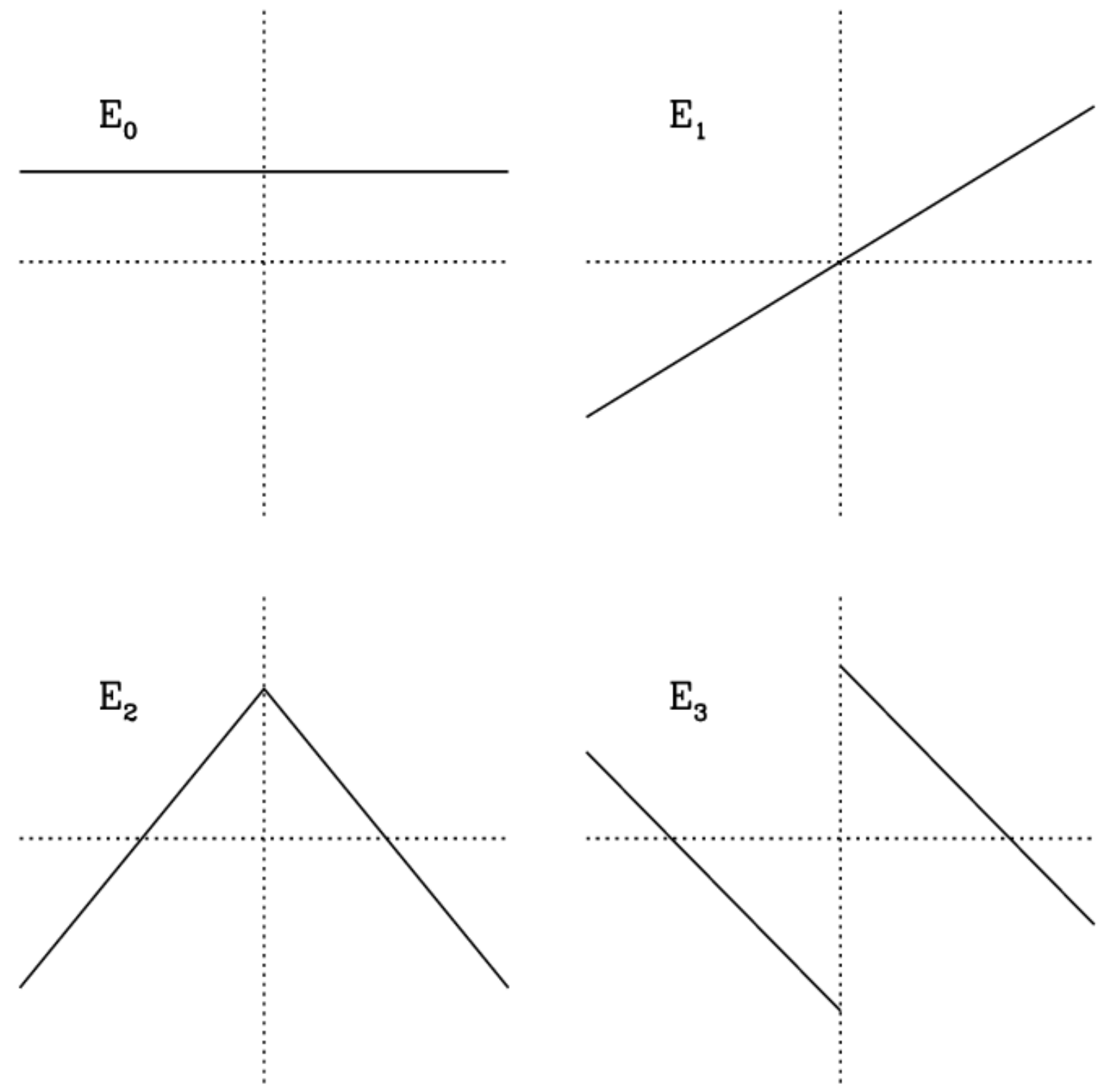}}
\caption{ The functions defined by
equations~\ref{eezero}-\ref{eethree}, where $u$ is plotted on the
horizontal axis. These four one dimensional functions can be used to
construct a set of S$_8$ octree basis functions as explained in the main
text.}
\label{seight-elements}
\end{figure}

The set of four $E_i$ functions consists of the pair with the
functional forms of the $P_0$ and $P_1$ Legendre polynomials plus a
pair of functions that have discontinuities about the origin.  The
left and right halves of this latter pair are linear combinations of
$P_0$ and $P_1$ Legendre polynomials, scaled to half the width, and
one is even and one odd about the origin.  The four functions are:
\begin{equation}
  E_0(u) = \begin{cases}   
                         \phantom{-}1 &  \text{if $ -1\leq u < 1$;} \\
                         \phantom{-}0 &  \text{otherwise.}
         \end{cases}\label{eezero}
\end{equation}
\begin{equation}
  E_1(u) = \begin{cases}   
                         \sqrt{3}u &  \text{if $ -1\leq u < 1$;} \\
                         \phantom{-}0 &  \text{otherwise.}
         \end{cases}\label{eeone}
\end{equation}
\begin{equation}
  E_2(u) = \begin{cases}  \sqrt{3}(1-2u) &  \text{if $ 0\leq u < 1$;} \\
                          \sqrt{3}(1+2u) &  \text{if $ -1 \leq u < 0$;}  \\
                         \phantom{-}0 &  \text{otherwise.}
         \end{cases}\label{eetwo}
\end{equation}
\begin{equation}
  E_3(u) = \begin{cases} \phantom{-}(2-3u) &  \text{if $ 0\leq u < 1$;} \\
                        -(2+3u) &  \text{if $ -1 \leq u < 0$;}  \\
                         \phantom{-}0 &  \text{otherwise.}
         \end{cases}\label{eethree}
\end{equation}
 These functions are shown in Figure~\ref{seight-elements}. With
respect to the origin the $E_0$ and $E_2$ are even functions and $E_1$
and $E_3$ are odd functions.  By construction all four functions are
orthogonal and obey a normalisation condition:
\begin{equation}
    \int_{-1}^1 E_i(u)E_j(u){\rm d}u = 2\delta_{ij}.
\end{equation}

 We combine these functions to create an analogous set
of three dimensional functions to eqn~\ref{sone-funcs} to give:
\begin{equation}
  G_{ijk}^l({\bmath x}) =  \frac{1}{\Delta^{3/2}_l} E_i\left(\frac{2x_1}{\Delta_l}\right)
   E_j\left(\frac{2x_2}{\Delta_l}\right) E_k\left(\frac{2x_3}{\Delta_l}\right),
\label{seight-funcs}\end{equation}
 where $i$, $j$, $k$ are the integers either zero, one, two or three,
and $x_1$, $x_2$, $x_3$ are the Cartesian components of ${\bmath x}$
and it is assumed the origin is the centre of the octree cell at level
$l$. The functions are mutually orthogonal and obey a normalisation
condition:
\begin{equation}
 \int G^l_{ijk}({\bmath x}) G^l_{lmn}({\bmath x}) {\rm d}^3{\bmath x} = \delta_{il}
   \delta_{jm}\delta_{kn},
\end{equation}
where the integral is over the volume of the level $l$ octree cell.
 There are 64 functions in total.  The eight functions defined by the
 values of $i$, $j$ and $k$ all being zero or one, are the S$_8$
 Legendre block functions at level $l$.  The functional forms of the
 remaining 56 functions are those of the S$_8$ octree basis functions.
 All of the basis functions have a least one discontinuity in value or
 slope about the principal coordinate planes defined by $x_1=0$, $x_2
 = 0$ and $x_3=0$. All 64 functions can be built from combinations of
 S$_8$ Legendre blocks placed in the eight level $l+1$ child cells.

 For practical applications it is easier to work with the smaller set
of eight Legendre block functions rather than the 56 distinct octree
basis functions.  The actual S$_8$ basis functions used for \WNF\ are
defined in terms of Legendre blocks in Appendix~\ref{OCT-DET}.

\section{Choosing a suitable pseudorandom number generator\label{FIND-GEN}}
  Once a set of octree basis functions has been chosen, the next step
to creating a realisation of a Gaussian white noise field is to assign
a value drawn from a Gaussian probability distribution to each octree
basis function.  This requires choosing a pseudorandom number
generator and establishing a mapping between the linear pseudorandom
sequence it produces and the octree basis functions.  We will discuss
the mapping first as the requirements of the mapping drive the choice
of pseudorandom generator.

  The octree functions form a four dimensional discrete space. For a
given choice of building block functions there will be a fixed number of octree
functions per octree cell.  We can develop a mapping as follows: 
\begin{itemize} 
\item{} 
firstly establish an ordering of the different types of octree basis functions
belonging to each octree cell; 
\item{} secondly an ordering of the octree cells at
a given level of the octree using a raster scan pattern over the three
physical dimensions of the octree;
\item{} finally an ordering by octree level
starting at the root node and descending.
\end{itemize}
  We give the full details of
the ordering used for \WNF\ in Appendix~\ref{MAP-OCT}.

 In general, a randomly chosen point in the root cell will overlap an
infinite number of basis functions. The values of the expansion
coefficients of these functions are determined by an infinite series
of short segments of the pseudorandom sequence which are spaced
progressively further and further apart as the octree is descended.
If the cost of accessing these coefficients were proportional to the
linear separations between these segments it would be impossible in
practice to descend far from the root cell. This would be a major
limitation for the method.

 To avoid this limitation we need a generator that allows essentially
random access to the entire sequence at a reasonable computational
cost. The ability to jump $N$ places in at worst of order $\log N$
time is highly desirable as this opens up the possibility of using the
entire period of the generator.

  There are ways of accessing a pseudorandom sequence that are
independent of the jump size. This can be done using an encryption
algorithm which takes as input the linear position on the pseudorandom
sequence and then scrambles the position in a highly nonlinear way to
produce a pseudorandom number. This kind of calculation is typically
quite expensive if the aim is to produce cryptographically strong
pseudorandom numbers.  However, it is possible to produce a relatively
fast pseudorandom number generator with this approach, albeit one that
should probably not be used for encryption.  The {\sc ran4} random
number generator from \cite{Numerical_recipes92} is an example of this
type. The authors found the {\sc ran4} generator produced good quality
random numbers for sequences of a billion numbers.  While we used this
generator for a prototype to \WNF, this particular generator is not
suitable for our purposes as the total sequence length is too
short. In principle it ought to be possible to devise a generator
along similar lines to {\sc ran4} but with a longer sequence.
 
 Rather than take this route we decided instead to take a generator
that has been described in the literature and that has been in common
use and which has been well tested.  Following a 
recommendation\footnote{The author is grateful to Stephen Booth of the Edinburgh
Parallel Computing Centre for suggesting a suitable generator and
providing me his own f90 implementation of the generator}, we have used a
generator first published in \cite{Lecuyer93}. The generator has
several names in the literature, but we will call it \MRG.  This
generator is available as part of the GNU scientific
library\footnote{http://www.gnu.org/software/gsl/} where it is called
{\sc gsl\_rng\_mrg}.

 The internal state of the \MRG\ can be represented as a five element
column vector with each element being an integer in the 
inclusive range $0,m-1$,
where $m$ is the prime $2^{31}-1 = 2,147,483,647$.  Given
the $n$-th state, $T^n_i$, the next internal state is generated by a
matrix operation: $T^{n+1}_i = M_{ij} T^n_j \mod m$, where modular
arithmetic, base $m$, is applied to the results of the matrix
multiplication.  A uniformly distributed pseudorandom number between
zero and one is
associated with each state:
\begin{equation}
   r_n =
 \begin{cases}
        \frac{T^n(1)-1/2}{m} & \text{if $0 < T^n(1) < m$}; \\
                             &  \\
        \frac{ m-1/2}{m}     & \text{if $ T^n(1) = 0$.} \\
  \end{cases}
\label{def_rand}\end{equation}
The matrix for advancing one step for \MRG\ is:
\begin{equation}\label{template}
 M_{ij} =\scriptstyle
 \begin{pmatrix}
      a_1 & 0 & 0 & 0 & a_2 \\
         1 & 0 & 0 & 0 & 0 \\
         0 & 1 & 0 & 0 & 0 \\
         0 & 0 & 1 & 0 & 0 \\
         0 & 0 & 0 & 1 & 0 \\
 \end{pmatrix},
\end{equation} 
where $a_1 = 107,374,182$ and $a_2 = 104,480$. The state vector
for the starting point of the sequence used for \WNF\ together with
some a few other examples of state vectors are given in
 Table~\ref{list_of_states}.

 Although cast as a matrix multiplication above, the operations required
to advance the generator one step can be implemented very efficiently
on a computer.  Because matrix multiplication is associative and
the property of associativity is not affected by the application of
modular arithmetic base $m$ to matrix multiplication, the jump matrix
which advances the state $N$ steps at a time is just $M^N\mod m$.  The
use of modular arithmetic ensures that the 25 coefficients of this
matrix always remain in the range $0,m-1$.  The cost of advancing the
state vector with a general matrix of this form is typically thirty
times more expensive than advancing by a single step. The costs of
building a jump operator, $M^N\mod m$, starting from
eqn~\ref{template} is $O(\ln N)$ which makes it practicable to
assign virtually the whole pseudorandom sequence to the octree basis
functions.

 From a computational point of view the cost of evaluating the
properties of a single octree cell in \WNF\ is still very expensive as
it requires the evaluation of typically thousands of pseudorandom
numbers at many different places in the sequence.  However the access pattern
needed for making initial conditions is highly localised so the actual
number of pseudorandom numbers that need to be evaluated per cell is
close to just ten. Using a raster scan access pattern for the cells
which mirrors the mapping of the pseudorandom sequence onto the basis
functions minimises the number of large jumps required and maximises the
number of consecutive accesses to the sequence. Further improvements
in speed can be obtained by caching the results of previous
evaluations.  With current processors it typically takes about $2\mu$s
on average per cell to return the expansion coefficients belonging
to that cell. There is still scope to improve the speed of the code
with further optimisations, but in practice the generation of the
white noise field typically takes only about 20\% of the time to
generate initial conditions, excluding the I/O.

 As stated in the documentation for the GNU scientific library the
full period of the \MRG\ generator is $P_{\rm gen} = m^5-1 \simeq
4\times10^{46}$. This means that the period consists of every possible
state vector with the exception of the null vector.  Interestingly the
generator has a sub-period, $P_{\rm sub} = (m^5-1)/(m-1) \simeq
2\times10^{37}$. Over multiples of this sub-period the jump matrix
becomes a multiple of the identity matrix (Stephen Booth private
communication). When this multiple (which ranges from 1 to $m-1$)
takes on small values, there are significant correlations between
pseudorandom numbers at this precise separation.  The large size of
the sub-period precludes any possibility of such a coincidence
occurring for levels shallower than 40 in the octree. Going deeper still
in the tree there is a remote possibility that some pseudorandom
numbers separated by a multiple of $P_{\rm sub}$ may occur in a given
set of initial conditions.  The chances of this happening, however, are
vanishingly small for any randomly chosen region.

  Of more concern is the more general question of whether this generator
provides sufficiently good pseudorandom numbers for making
cosmological simulations. It is desirable that the generator passes a
diverse set of randomness tests. However, even in the epoch of
precision cosmology the requirements on a generator for cosmological
simulations are less strict that many other applications such
cryptography. Some deviation from randomness are acceptable for
cosmological initial conditions provided they are sufficiently
small. The fact that the pseudorandom numbers are discrete and not
truly uniformly distributed is not a concern, although as described in
Appendix~\ref{MAP-OCT} we do take measures to mitigate the
discreteness effect to ensure that the tail of the Gaussian
distribution for the Gaussian pseudorandom numbers is properly
populated.  

 For some purposes such as encrypting secret messages or for gambling
machines it is highly desirable that the pseudorandom sequence cannot
easily be predicted by studying a small part of its output sequence.
In this respect \MRG\ performs poorly as just five consecutive numbers
are sufficient to deduce all five elements of the state vector.  Once
the state vector is known the whole sequence is determined.  This
property means that an $n$-tuple (for $n>5$) of pseudorandom numbers
produced by \MRG\ cannot uniformly sample the $n$-point joint
probability function.  However, the coefficients, $a_1$ and $a_2$, used
in the generator were selected in part by a requirement than the
deviations from uniformity in the $n$-point joint probability function
for $6\le n<21$ are confined to very fine scales \citep{Lecuyer93}. In
practice this means that detecting deviations from randomness in the
joint $n$-point function requires very large samples in order to
distinguish non-uniformity from shot noise. This small scale
non-uniformity is not obviously a problem for cosmological initial
conditions as any effect is likely to be dwarfed by sources of
numerical error in any actual N-body simulation.

 A second feature of \MRG\ is that over the entire period, with one
exception, each pseudorandom number appears $m^4$ times exactly - a
highly subrandom pattern.  So a test looking at the frequencies of
occurrence of different pseudorandom numbers will show a failure for a
sufficiently large sample of the sequence.  This deviation from
randomness is extremely small so hardly a concern for making initial
conditions.  Given these known limitations it is interesting to see
how well \MRG\ does perform in wide range of standard tests of
randomness.
   
The authors of \cite{Lec_simard2007} have developed a software
library, {\sc testu01}, for testing pseudorandom number generators
empirically. The results of libraries of tests performed on many
common generators, including \MRG, are given in this paper.  Their
most rigorous test battery, called BigCrush, yields 160 independent
statistical results from a total of 106 separate tests.  A total of
$2.7\times10^{11}$ pseudorandom numbers for each generator are used in
these tests. The outcome of each test is a p-value, which for a
perfect generator would be expected to be drawn from a uniform
distribution between 0 and 1.  A generator is said to fail a test if
the p-value is within $10^{-10}$ or zero or unity. A test where the
p-value is within $10^{-4}$ of the same limits is noted as suspect.
Quite a few commonly used generators do fail multiple tests.  The
\MRG\ generator does not fail any of the BigCrush tests and none of
the p-values are suspect.

This is encouraging and on this basis we are content to use this
generator. The size of the sample tested is larger than would
typically be used to generate most sets of initial conditions, although
some like those of the \MXXL\ require more.  We do nonetheless expect
this generator to eventually fail a randomness test applied to a
larger sequence of numbers than tested by BigCrush for reasons
explained above. The deviations from randomness we described are small
and not a great concern.  Tests of pseudorandom number generators,
however, can never completely set the mind at rest as it is impossible
to be sure that some new random test may reveal a significant flaw
previously undetected.

\section{Adding Panphasia to the \ICgen\ code\label{ADD-ICGEN}}
 In this section we describe how to modify the \ICgen\ code for making
cosmological initial conditions to use \WNF\ to set the phase
information.  There are two main goals to this section.  Firstly to
show that it is possible to make initial conditions that accurately
reconstruct the phases defined by \WNF\, and secondly to act as a
practical guide to help anyone wanting to add \WNF\ phases to other
initial condition generators.

\subsection{Overview of \ICgen.}
 The \ICgen\ code is used by the Virgo consortium to generate Gaussian
initial conditions for a variety of projects relating to large-scale
structure, galaxy formation, the internal structure of dark matter
haloes and the formation of the first stars. The code is able to make
resimulation initial conditions with displacements and velocities
calculated using second-order Lagrangian perturbation theory ({\sc
2lpt}). While the initial conditions we make for this paper use this
feature, there is no significant interaction between the methods we
describe here and those required to generate {\sc 2lpt} initial
conditions. We will describe only the most relevant features of the
code here.  A more detailed description of the code, including the
method to make {\sc 2lpt} multi-scale initial conditions can be found
in \cite{Jenkins2010}.

 When it comes to modifying the code to use the phase information from
\WNF\ it is useful to divide the kinds of initial condition that \ICgen\ can
make into two classes.  The first class are cosmological
initial conditions where a single Fourier grid is used to
generate displacement and velocity fields for all the particles in a
large periodic domain. An example of this is the \MXXL\ simulation
\citep{Angulo2012}, which modelled 303 billion particles in a cubic
volume of about 4 Gpc on a side.

The second class, which we will call resimulation initial conditions
(or equivalently zoom simulations) uses multiple Fourier grids to
compute the displacement and velocity fields to generate multi-scale
initial conditions.  In common with cosmological initial conditions a
grid is used that covers the entire simulation domain. We will call
this the parent grid or outer grid.  The extra grids, which we will
call collectively the inner grids,  typically have a
similar number of grid points as the parent grid, but are physically
smaller and placed concentrically around a focal point of interest in
the simulation volume.  The \ICgen\ code can place many nested
sub-grids around a point allowing it to make very high resolution
initial conditions for the region that is contained within all of the
grids.  The remainder of the simulation volume is required to provide
the appropriate tidal forces only and is modelled at lower mass
resolution to reduce computational cost.  A recent example of
resimulation initial conditions made by \ICgen\ are those created for
the Phoenix project \citep{Gao2012}, the computational goal of which
was to model at high numerical resolution the dark matter in
individual galaxy clusters, selected from the Millennium simulation
\citep{Springel05}.

 The \ICgen\ code uses Fourier methods to generate the Gaussian
displacement and velocity fields for each grid.  The fluctuations are
created in $k$-space by generating independent random amplitudes and
phases for each Fourier mode (subject to the condition the field is
real) with the option of using one of several different pseudorandom
number generators to calculate a series of independent Gaussian
pseudorandom numbers.  The amplitudes of the Fourier modes are scaled
appropriately so as to reproduce the desired linear density fluctuation
power spectrum.  For a given grid the Fourier modes are set
only in a range between low-$k$ and high-$k$ limits.

 For cosmological initial conditions utilising a single grid these
limits in $k$ are determined by the fundamental mode of the simulation
cube at low-$k$ and typically by the particle Nyquist frequency at
high-$k$. The high-$k$ cut-off that is chosen to be spherical so that the
fluctuations are isotropic at small scales.  For resimulation initial
conditions the low-$k$ and high-$k$ limits associated with each nested
grids in real space are dovetailed to ensure that the power spectrum
in the high resolution region has the appropriate contributions all
the way from the fundamental mode of the simulation cube down to the
particle Nyquist frequency of the high resolution region.

  All grids are treated in the same way by \ICgen\ which means that all
field quantities are periodic on the physical scale of the grid.  This
periodicity is only strictly correct for the parent grid. However the
affect of periodicity on the other grids can be limited by choosing
a low-$k$ cut-off is significantly larger than the fundamental mode of
that grid.  If this condition is met then the correlation length of
the field on the grid is much smaller than the size of the grid itself.

  Once the displacement and velocity fields (and other fields) have
been calculated on a grid the \ICgen\ code uses interpolation to
compute the values of these fields at the locations of unperturbed
particles. The \ICgen\ code uses the value of the fields and their
spatial derivatives in the interpolation from the grid points to the
positions of the particles in the unperturbed particle load. The
interpolation scheme used is described in detail in the appendix of
\cite{Jenkins2010}. 

  For a resimulation to be successful the phases present in the
 original cosmological simulation must be reproduced on the parent
 grid. For a resimulation with higher resolution in some region of
 interest than the parent simulation additional high-$k$ power is
 needed.  The choice of the phases for this extra power is not
 constrained by the large scale power.  In \ICgen\ the phase
 information for the high-$k$ power is determined by a series of
 arbitrarily chosen pseudorandom number seeds: one for each extra
 grid. The precise positioning and dimensions of the extra grids is
 ill determined: all that is required is that it be approximately
 concentric about the region of interest.  All of these freedoms
 contribute to making the choice of the phase information at high-$k$
 power rather arbitrary.  In practice the phase information is encoded
 in a text parameter file the length of which increases as the number
 of grids increases.

\subsection{Adding cosmological initial conditions from Panphasia to \ICgen.\label{cosmo_ics}}

  The phase information for cosmological initial conditions is most
compactly represented as a finite set of amplitudes and phases each
associated with a periodic plane wave that spans the simulation
volume.  These waves range in wavelength from the fundamental modes of
the periodic simulation volume to a cut-off wavelength typically
determined by the interparticle spacing. This contrasts with the
equivalent octree basis function representation, where in order to
reproduce the phases of these particular waves exactly, an infinite
number of expansion coefficients are needed. Nonetheless it is
possible to accurately reproduce the phase information with a finite
number octree basis functions, particularly when using the S$_8$
octree functions introduced in Section~\ref{OCT-GEN}.  In practice
only octree basis functions down to some maximum depth, $l_{\rm max}$,
in the octree can be used, where the value of $l_{\rm max}$ should be
as large as possible given limited resources.

  The \ICgen\ code makes initial conditions using Fourier methods
starting with a $k$-space representation of a Gaussian field on a cubic
grid.  The natural way to incorporate \WNF\ is to choose a Fourier
grid which is commensurate with the Legendre block expansion of \WNF\
at level $l_{\rm max}$ of the octree.

 We assume that a particular cubic region within \WNF\ consisting of
$N^3$ whole octree cells at level $l$ has been selected to define the
phase information for a given cosmological simulation. The corresponding dimension of the cubic Fourier
grid, $M$, should obey $M = 2^{(l_{\rm max}-l)}N$, where $l_{\rm
max}\ge l$.  This ensures that there is a one-to-one correspondence
between grid points and octree cells at level $l_{\rm max}$.  The use
of fast Fourier transform algorithms places restrictions on the value
of $N$, limiting it to be a product of small prime factors only.

 Having matched the grid to \WNF\ the \ICgen\ code takes only the
information provided by \WNF\ as deep as level $l_{\rm max}$. 
  For each octree cell at level $l_{\rm max}$ we know the basis function
coefficients of the expansion of \WNF\ for S$_8$ Legendre blocks.  We
can associate a separate grid to each of the eight types of Legendre
block, and assign an expansion coefficient to a corresponding grid
point for all cells. Each of these grids is an independent discrete
realisation of a Gaussian white noise field. The remaining task is to
combine the information on these eight fields to produce a single
field on a grid that accurately reconstructs the \WNF\ phases.

Taking the grid points to represent delta-functions, scaled by the
corresponding values of the expansion coefficient, we can in principle
exactly regenerate \WNF\ truncated to level $l_{\rm max}$ by convolving each
of the eight grids with the appropriate Legendre block and coadding
the results to give a single continuous field. This continuous field
would nonetheless have a discrete representation in $k$-space because
of periodic boundary conditions.

 In practice in \ICgen\ the eight grids are combined in $k$-space to
give a discrete combined field. This is done by applying a fast
Fourier transform to each real grid to produce a $k$-space
equivalent. Once in $k$-space the convolution with the appropriate
Legendre block is achieved by multiplying by the Fourier transform of
the Legendre block (given by eqn~\ref{sph_bess_fn_def}).  An
additional phase factor corresponding to a uniform translation in
real-space, has to be included in this convolution for \ICgen.  This
is because \ICgen\ places a grid point at the coordinate origin, which
means that the grid points and the octree cell centres are everywhere
displaced from each other by half a grid spacing in each of the
Cartesian directions. The translation by half a grid spacing in all
three Cartesian directions is need to ensure that the phase pattern
appears in the correct physical location.  Summing the eight fields
produced by the convolution results in a discrete and bandwidth
limited representation of \WNF\ in k-space. This field is not a true
Gaussian white noise field as the S$_8$ Legendre blocks placed at a
given level of the octree are not a complete basis set.

 The field produced this way can be restored to a true white noise
field by adding an additional uncorrelated field with an ensemble
averaged power spectrum:
\begin{equation}
P_{\rm add}^l({\bf k})  = 1-\sum\limits_{{\rm S}_8} j^2_{i_1i_2i_3}({\bf k}\Delta_l).
\label{add_level_power}\end{equation}  
Using the identity eqn~\ref{sph_bessel_identity} 
it is easy to see that the power spectrum of the truncated field
averaged over all directions deviates as $k^4$ from a white noise field
at large spatial scales for the Legendre block functions of S$_8$, and
as $k^2$ for the S$_1$ Legendre block.

 This extra component should ideally be generated from the basis
expansion coefficients of \WNF\ for levels deeper than $l_{\rm max}$, but
there has to be a cut-off in practice and we take it to be $l_{\rm max}$.  So
instead \ICgen\ generates a ninth independent white noise grid, and
uses the form of $P_{\rm add}({\bf k})$ so that when combined with the
other eight fields the result is a true white noise field.  This
guarantees that the initial conditions have the correct power
spectrum with an isotropic cut-off in $k$, but comes at the price
that the phases of \WNF\ particularly at small scales are not
perfectly reconstructed.  The pseudorandom numbers needed for the
ninth grid are not part of \WNF, but as explained in
Appendix~\ref{MAP-OCT} the coefficients for the ninth field are 
generated at the same time as the basis coefficients for the other
eight grids.

 The requirement to build the field with so many components
potentially has some negative practical effects on the memory
efficiency of the code.  The memory requirements would be
significantly increased if all nine grids had to be stored at the same
time when using the coefficients for all eight Legendre blocks.  In
fact the code already needs to be able to store four grids in order to
make 2lpt resimulation initial conditions as described in
\cite{Jenkins2010}.  These grids however are not required until the
white noise field has been computed.  So there does not need to be any
increase in the memory usage of the code provided the white noise
field is evaluated three times in succession.  In practice, the actual code stores
five grids and therefore has to evaluate the white noise field
twice. As it takes both more cpu time and more memory to make initial
conditions using S$_8$ in preference to S$_1$, there has to be a good
reason to prefer it.   This applies even more strongly to $S_{27}$
which would require 28 grids to be combined. To keep the memory
requirements the same with $S_{27}$ would require evaluating the 
white noise field six times, and each evaluation of the white
noise field would be significantly more expensive too. There would
need to be an even stronger reason for rejecting S$_8$ before considering
the  $S_{27}$ set.

 The ultimate decision as to the best set of octree basis functions is
made later in the paper from studying the end states of simulations.
It is useful nonetheless as a guide to help interpret the results of
these simulations to compare the different choices of octree basis
function by looking at linear density fields first.  We define first
an error field, which is the difference between the linear density
field reproduced by the method, and the true \WNF\ linear density
field.  We can then characterise this error field by its power spectrum.
Because the ninth field is independent of \WNF\ the error power
spectrum is simply $2P_{add}$.

Using this error power spectrum we can quantify the effect of this
error field on density fluctuations in the initial conditions, for
some assumed power spectrum $P(k)$, by determining the fractional
error in the RMS fluctuations as a function of volume: $\epsilon_{\rm
RMS}$ as:
\begin{equation}
 \epsilon_{\rm RMS}^2 = \frac{ \int 2P_{\rm add}  k^2 W^2(k){\rm d}k}
                           { \int P(k) k^2 W^2(k){\rm d}k}.
\label{error_field}
\end{equation} 
The top and bottom terms on the r.h.s. are the RMS fluctuations in the
error field and the true field respectively, smoothed by a suitable
spherical filter, $W$, with some characteristic scale that can be
varied. For spherical perturbations the collapse epoch is determined
by the linear overdensity at some fiducial epoch, so these fraction
RMS fluctuations give an indication of how structure formation is
affected as a function of scale.

\begin{figure}
\resizebox{\hsize}{!}{
\includegraphics{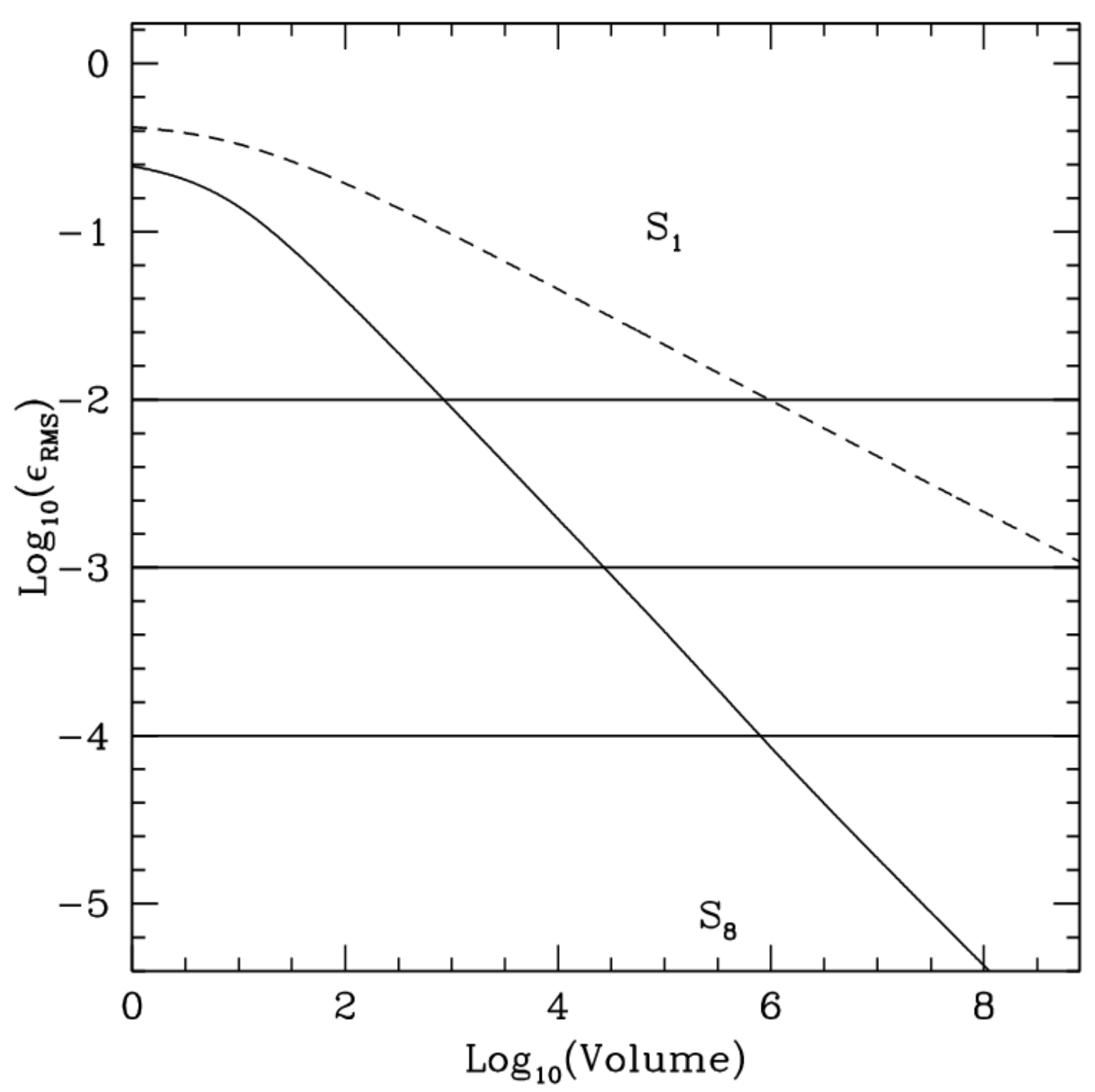}}
\caption{ A comparison of the relative accuracies
in reproducing the correct phases in initial conditions
for two choices of sets of Legendre blocks. The quantity
of the $y$-axis, defined in eqn~\ref{error_field}, is plotted
as a function of the volume of the smoothing filter in
units of the volume of the grid cells. The integrals have
been truncated at the grid Nyquist frequency. }
\label{frac_gauss}
\end{figure}

  In Figure~\ref{frac_gauss} we plot $\epsilon_{\rm RMS}$ against the
effective volume of the filter, $W$, for power-law initial conditions
with $P(k)\propto k^{-2.75}$ made with either the S$_1$  or S$_8$ Legendre
blocks.  We take $W$ to be a Gaussian filter with the zeroth and
second moments matched to a spherical tophat with a volume plotted on
the $x$-axis in units of the volume per grid cell.  The tophat filter
itself is unsuitable because the top integral in eqn~\ref{error_field}
is dominated by a surface term rather than by the volume term for the error
field generated by S$_8$.  The choice of power-law index is appropriate
for \CDM\ initial conditions with fluctuations populated down to a
particle Nyquist frequency corresponding to a particle mass of about
$10^6$~\msun which will feature in the tests later in this section.

Clearly the field generated by S$_8$ is a much better approximation to
\WNF\ for all volumes than S$_1$, with the difference increasing with
increasing volume.  For a fractional error of $\epsilon_{\rm
RMS}=0.01$ there is a factor of about three orders of magnitude in volume
between the S$_1$ and S$_8$ initial conditions.  This means that to
achieve the same accuracy in the phase reconstruction using initial
conditions made using the S$_1$ octree basis functions requires orders
of magnitude higher numerical resolution than for S$_8$.  In
Section~\ref{SIM-TESTS} we will see how these differences translate to
the end states of simulations. Before this however we now
consider the task of how to adapt \ICgen\ to make resimulation initial
conditions.

\subsection{Adapting \ICgen\ to make resimulation or zoom simulations using \WNF.\label{resim_method}}
As described earlier the \ICgen\ code generates what we will call
Fourier resimulation initial conditions in a piecewise fashion by
calculating the displacement fields on a series of nested grids about
some focal point of interest in the parent simulation. The total
linear displacement for any particle is just the sum of the linear
displacements generated by the individual grids that it spatially
overlaps. The displacement fields for each grid are generated in the
same way as for cosmological initial conditions with starting point
being to generate a $k$-space representation of a Gaussian field on a
grid.  To adapt \ICgen\ to use \WNF\ requires finding a way to
generate equivalent $k$-space Gaussian fields based on \WNF. Once this has
been achieved no further modification of the code is needed to produce
initial conditions.
 
 The precise placement of the Fourier grids in \ICgen\ Fourier
resimulation initial conditions is largely arbitrary. All that is
required is that the different grids be nested and centred
approximately on the region of interest. With the Fourier method the
displacement field pattern is tied to the grid and would move rigidly
with the grid if it was decided to place the grid in a slightly
different location.  In contrast the octree basis functions of \WNF\
have fixed coordinates so moving the grid will not lead to the phase
pattern itself shifting.  In practical terms this means that it is not
necessary when using \WNFIAN\ phases to publish the grid positions.
It is true that the 'error fields', that is the differences between
the true initial conditions and those actually generated will depend
on the choice of grid positions, but as the aim in numerical work is
to minimise the numerical effects to the point where they do
not affect the scientific conclusions it should not be necessary to
specify the precise grid positions.

 As with cosmological initial conditions the extra grids must be
arranged so that the grids are commensurate with the octree cells
which means there is only a discrete set of positions where the grids
can be placed.  These positions are those where the grid points are
located at the corners of octree cells.  The sizes of the grids
measured in grid cells are also restricted for computational 
and numerical reasons, by both the method of parallelisation
in \ICgen\ and through the use of fast Fourier transforms.
In practice the ratio of the physical sizes of the nested
grids have to be simple fractions, and for all the tests in
this paper they differ by powers of two only.

 In the Fourier method for making resimulation initial conditions the
phase information on each grid is independent from that on every other
grid simply because each Fourier mode is set with a different
pseudorandom number and each Fourier mode belongs to a single grid.
To make the grids independent using \WNF\ all that needs to be done is
to assign each octree basis function to a single grid. Any alternate
approach that allows an octree basis function to be split between
grids does not appear attractive as the contributions on the different
grids would be sampled on different scales and, because they are
coherent, add in a complex way with associated fringing.

While the octree basis functions themselves are perfectly confined to
a single cell, once a convolution has been applied to generate the
initial conditions, the information contained within a single cell is
propagated to all points on the grid, although in a far from uniform
way.  As discussed in Appendix~\ref{OCT-DET} both the zeroth and
first moments evaluated about the cell centre of the octree basis
functions are identically zero.  This vanishing of the zeroth and
first moments means that the information stored in the octree basis
function is more strongly localised than for the white noise as a
whole occupying the same volume. This property of partial locality
makes the size of edge effects when making multi-scale initial
conditions using Fourier methods smaller than might first be
supposed. 

 As the octree basis functions are built out of Legendre blocks the
actual mechanics of calculating the $k$-space Gaussian field for
resimulation initial conditions is the same as for cosmological
initial conditions.  The only extra ingredient needed to make \WNF\
resimulation initial conditions is to decide how to partition the
octree basis functions between a given set of nested grids. While the
majority of octree basis functions only overlap the outer most grid so
there is no choice, for all other octree functions there is a choice
between two or more grids. The choice should ideally be that that
gives the best initial conditions, although it is hard to be precise
about exactly what best means in this context.  We have opted to adopt
a heuristic approach to the solution of this problem.  The
justification of this approach is that it can be demonstrated to work
well in practice as we show later.  This heuristic approach can be
described by four rules.

\begin{itemize}

\item{Rule 1: } {\it `Any octree basis function that can be included,
should be included'}.  As every octree basis function contributes to
the phase information, missing out any of the octree basis functions
will degrade the quality of the phase reconstruction. This rule makes
it easy to count how many basis functions must be used in total over
all grids and makes it simple to check in the code that all have been
placed.

\item{Rule 2: } {\it `With the exception of the outermost grid, only
 whole octree basis functions can be assigned to a grid.'}  The reason
 for this rule is to minimise unwanted edge effects. As explained
 earlier in this section in \ICgen\ all of the grids are periodic.
 These are the correct boundary conditions only for outermost grid,
 and it is therefore necessary for all other grids to try and minimise
 the edge effects due to periodicity. Only whole octree basis
 functions are guaranteed to have vanishing zeroth and first moments
 and this is required to limit their range of influence as discussed
 earlier in this subsection. The requirement that whole octree basis
 functions are placed in these grids puts further restrictions on the
 possible sizes and precise placements of the inner grids.  The
 smallest octree basis function measures two cells along each edge of
 the grid, which means both that the grid dimensions must be even, and
 the edges of the grids must line up with the octree cells these basis
 functions occupy.

\item{Rule 3: } {\it `Except where rule 4 is broken, an octree basis
 function should be placed on the innermost grid allowed by rule 2.'}
 Far from any boundary it is desirable that an octree basis function is
 represented over as many grid cells as possible.  This is because the
 Fourier transform of the octree basis functions is not intrinsically
 bandwidth limited but its representation on a grid will be  bandwidth
 limited. Placing a basis function over as many cells as possible minimises
 the truncation of power at small scales.  Near a grid boundary however
 there is another factor to consider which requires rule 4.

\item{Rule 4:}{\it `For grids other than the outermost, and for octree
   basis functions other than the smallest, no octree basis function
   shall be placed within a perpendicular distance, measured from its
   edges, from the grid boundary that is less than a factor X times
   its own edge size.'}  A rule of this kind is required to minimise
   unwanted edge effects due to periodicity for all but the outermost
   grid.  The reason for not wanting to place large octree base
   functions close to a grid edge is simply that the larger an octree
   basis function is the further its influence spreads in the initial
   conditions.  This requirement is in direct opposition to rule 3
   which encourages larger octree cells where possible. The use of a
   distance criterion with a factor `X' to be determined empirically
   ensures a compromise is possible. For scale-free initial conditions
   it would be natural to take `X' to be constant in the absence of
   any characteristic scale other than the cell size itself. For
   \CDM\ models, for example, the power-law index varies with scale but at
   small scales is slowly varying. As there is a smallest octree basis
   function that can be placed on a grid, and rule 1 requires these
   smallest basis functions to be present, they have to be treated
   exceptionally and are allowed to be placed right up to the
   boundaries.
 \end{itemize}

  Once the factor X in rule 4 is decided there is then a unique
solution for placing the octree cells on a given set of grids.
Different values of X do in some cases result in the same placement.
Small values of X mean that the octree basis functions close to the
edges of the inner grids are large and therefore have longer range
effects.  Large values mean that octree basis functions are assigned
to small numbers of grid points and are therefore less accurately
represented.  As part of the next section we will compare dark matter haloes
generated with different values of X to determine a close to optimal
value.  A code which implements the rules above is included in
addition to \WNF\ in the public code release. See \companion.

\section{ Testing the accuracy with which the phase information
from Panphasia can be reproduced.\label{SIM-TESTS}} The main goal of this
section is to demonstrate that it is possible to make good quality
resimulation initial conditions using the phase information provided
by \WNF.  To do this we will test how well the methods for making
cosmological and resimulation initial conditions described in the
previous section succeed in this task.  We will judge success by
looking at the end states of a set of simulations at redshift zero.  A
further goal of this section is to provide a guide to help anyone
wanting to add \WNF\ to an initial conditions code for how to test the
code is actually working.

 Judging what constitutes `good quality' resimulation initial
conditions is inevitably rather subjective. There is relatively little
published about the efficacy or otherwise of resimulations that can
help define a standard. There is none at all that can be directly
compared with the results of this section.  This is because published
methods of setting the phases do not define the phase in an objective
way, so it is not possible to make the same initial conditions in two
different ways and expect them to be the same, which is what we aim to
do here.  Nonetheless it is reasonable to require that the size of the
errors we can determine in this section should be comparable or
ideally smaller than the sizes reported in the literature for typical
applications of resimulations. Hard numbers can be found in only a few
published papers.  These numbers show how well bulk properties of dark
matter haloes are reproduced for haloes resimulated at several
numerical resolutions.  As two simulations of the same halo at
different resolutions do not have identical phase information there is
no reason to expect that the bulk properties should be exactly
reproduced, but it is found for haloes represented by millions of
particles or more that properties such as virial mass or maximum
circular velocity are reproduced at sub-percent accuracies
\citep{Springel08,Hahn_11,Gao2012}.  We will use this crude measure to
judge whether the resimulation initial conditions produced from \WNF\
are of a comparable standard to published methods or not, and declare
them to be of good quality if they are.

 Ideally, we need a reference calculation that precisely reproduces
the phases from \WNF. We have seen from Subsection~\ref{cosmo_ics}
that this cannot be done, but it is possible in principle to produce a
very accurate approximation to the true phases by using an extremely
large Fourier transform to make cosmological initial conditions.  We
can then run a simulation using these initial conditions and create a
reference end state.  This can be compared to the end states of
simulations starting from cosmological and resimulation initial
conditions made using Fourier transforms of a size that would be used
in practice because they can be readily afforded.

 The plan for this subsection is as follows: in
Subsection~\ref{exp_sims} we introduce the reference calculation;
in Subsection~\ref{phase_measure} we describe and test a method to measure 
how well the phase information is reconstructed;
in Subsection~\ref{cosmo_test} we apply this measure to cosmological
initial conditions; in Subsection~\ref{resim-test} we demonstrate
that the proposed resimulation method works well; finally in
Subsection~\ref{resim_param_search} we investigate the sensitivity of
the method to changes in parameters such as the Fourier grid and the X
parameter introduced at the end of the last section.

\subsection{The simulations\label{exp_sims}}
 We have chosen a fairly typical case of the resimulation method
which is to resimulate a single isolated dark matter halo with a mass
similar to the inferred mass of the Milky way \citep{Springel08,
Stadel09}.  We have chosen a halo from a completed high resolution
N-body simulation run by the Virgo Consortium called \DOVE.  This
\LCDM\ dark matter only simulation is a 70.4~Mpc$/h$ periodic box with
similar mass resolution to the Millennium-II simulation
\citep{BK_09}. The cosmological parameters however differ from the
Millennium-II and are listed in Table~\ref{cosmo_params}. These
parameter values are taken from Table~1 of \cite{Komatsu11} and are based
on constraints derived from the \CMB, \BAO\ and the Hubble
constant. The \CDM\ transfer function for this model was calculated
using \CMBFAST\ \citep{CMBFAST}. The initial phases were taken from
\WNF\ and a $3072^3$ Fourier grid was used to make the initial
conditions. We give the precise location for the phases in \WNF\ in
Section~\ref{PUB-CONV}.  The simulation was run to redshift zero using
the \PGADGET\ N-body code \citep{Springel08}.

\begin{table} 
\begin{center}
\begin{tabular}{|l|r|}
\hline
      Cosmological parameter                       &            Value \\
\hline
   $\Omega_{\rm matter}(z=0)$  \phantom{xxxxxxxxx}  &  \phantom{xxxxxxxxx} 0.272\phantom{0} \\
   $\Omega_\Lambda(z=0)$       \phantom{xxxxxxxxx}  &  \phantom{xxxxxxxxx} 0.728\phantom{0} \\
   $\Omega_{\rm baryon}(z=0)$  \phantom{xxxxxxxxx}  &  \phantom{xxxxxxxxx} 0.0455           \\ 
   $H_0$~/km~s$^{-1}$~Mpc$^{-1}$\phantom{xxxxxxxxx}  &  \phantom{xxxxxxxxx} 70.4\phantom{000} \\
   $\sigma_8$                  \phantom{xxxxxxxxx}  &  \phantom{xxxxxxxxx} 0.81\phantom{00} \\
   $ n_s$                      \phantom{xxxxxxxxx}  &  \phantom{xxxxxxxxx} 0.967\phantom{0} \\
\hline
\end{tabular}
\end{center}
\caption{The cosmological parameters of the \DOVE\ simulation and for
all of the test simulations in this paper: $\Omega_{matter}$,
$\Omega_\Lambda$, and $\Omega_{\rm baryon}$ are the average densities
of matter, dark energy (with -1 equation of state) and baryonic matter
in the model in units of the critical density; $H_0$ is the Hubble
parameter; $\sigma_8$ is the square root of the linear variance of the
matter distribution when smoothed with a tophat filter of radius
8~\hmpc radius; and $n_s$ is the scalar power-law index of the power
spectrum of primordial adiabatic perturbations.}
\label{cosmo_params}
\end{table}

 The test halo was chosen by the author to have no close large
neighbours by visual inspection of dot plots.  Similar results to
those presented in this section have been obtained with a second MW
mass halo in a different part of the \DOVE\ volume. The resimulation
methods described in the last section have also been applied to
resimulate cluster mass dark matter haloes in other Virgo simulations
set up with \WNFIAN\ phases and the properties of these haloes are
reproduced to similar fractional accuracy as we report
for the halo studied in this section. We can therefore judge
the quality of the initial conditions by studying this one
typical halo.

 Although the \DOVE\ simulation was used to select a halo we will not 
use any data from the \DOVE\ simulation in this paper.  We have,
however, compared the properties of the halo for the highest quality
simulations in this paper with its counterpart in the \DOVE\ simulation
and find excellent agreement in its position and measured properties.

 We expect that the accuracy with which the phase information is
reproduced improves with the problem size as can be deduced from
Figure~\ref{frac_gauss}.  To test the methods we should therefore not
aim to resimulate a halo at very high resolution. Going to the other
extreme of low resolution would mean simulating a halo represented by
just a few particles which would make it difficult to determine any of
the halo properties due to the discreteness. As a compromise we have
chosen a resolution for the halo with sufficiently high resolution to
start to show the very rich substructure revealed by ultra-high
resolution simulations of dark matter halo formation
\citep{Springel08, Stadel09, Gao2012}.  The halo has about 250\;000
particles within $R_{200}$ and around fifty identifiable substructures
with more than twenty particles as determined by the \SUBFIND\ group
finder \citep{Springel_etal_01}.

\begin{table}
\begin{center}
\begin{tabular}{| l | l | c |} 
\hline
  Property     &  Unit  &   Value(s)        \\
\hline
  Centre           &  Mpc/$h$                & (43.1732,49.9170, 2.2070) \\  
  $M_{200}$          &  $10^{10}h^{-1}$~\msun  &   91.61  \\
  $M_{\rm vir}$      &  $10^{10}h^{-1}$~\msun  &   110.02  \\
  $R_{200}$          &  kpc/$h$                &  157.95   \\
  $R_{\rm vir}$      &  kpc/$h$                &  213.47   \\
  $V_{\rm disp}$     &  km~s$^{-1}$             &  102.72   \\
  $V_{\rm max}$      &  km~s$^{-1}$             &  181.71  \\
    $\lambda^\prime$   &     --                  &  0.0133   \\
  $(c1,c2,c3) $     &       --                & (-0.213, -0.651, -0.728)  \\
\hline                                  
\end{tabular}
\end{center}
\caption{Some properties of the reference halo at redshift zero. The
centre is the potential minimum. $M_{200}$ and $M_{\rm vir}$ are
masses within spheres centred on the potential minimum, with mean
densities of 200 and 97 times the critical density respectively.  The
radii of these spheres are $R_{200}$ and $R_{\rm vir}$. The velocity
dispersion, $V_{\rm disp}$ is that of the main subhalo as determined
by \SUBFIND. $V_{\rm max}$ is the maximum circular velocity, where all
circular velocities are calculated assuming spherical symmetry about
the halo centre.  The halo spin is given by $\lambda^\prime$, first
defined in {Bullock} {et~al.}, 2001.  The spin is determined for
all particles within a sphere of radius $R_{\rm vir}$ centred on the
halo potential, $\lambda^\prime = J/\sqrt{2}M_{\rm vir}R_{\rm
vir}V_{\rm vir}$, where $J$ is the angular momentum and $V_{\rm vir}$
is the circular velocity at the virial radius.  The direction of the
spin is given in terms of the directional cosines projected onto the
$(x_1,x_2,x_3)$ axes. }
\label{halo_props}
\end{table}

 We used the same software to build the particle load as the Aquarius
haloes \citep{Springel08}.  The lowest mass particles are placed in a
region that occupies a small fraction of the simulation volume.  The
particles making up the redshift zero halo and its immediate
surroundings are located within this high resolution region.  Higher
mass particles are placed further out around this region and provide
the appropriate tidal field on the region of interest.  The particle
load has 787\;939 high resolution particles each with a mass of
$6.24\times10^6h^{-1}$~\msun. The total mass within the \DOVE\ volume
is equivalent to about $1616^3$ of these particles. The smallest
Fourier transform that can be used to generate power down to the
Nyquist frequency of high resolution particles given the quantisation
constraints of \WNF\ is $3072^3$ -- the same grid size used to make the
\DOVE\ initial conditions.

 The displacements and velocities are calculated with  second
order Lagrangian perturbation theory using the method in
\cite{Jenkins2010} for all of the initial conditions.  The \PGADGET\
code was used to integrate the equations of motion from the start
redshift of 63 to redshift zero. The same numerical parameters for
\PGADGET\ were used for all simulations to ensure that any differences
observed are due only to differences in the initial conditions.  The
gravitational softening comoving length for the high resolution
particles was
$2h^{-1}$~kpc at all times.  The \SUBFIND\ group finder
\citep{Springel_etal_01}, which is integrated into \PGADGET. was run
on the high resolution particles only.  There are no heavier mass
particles near the main halo at redshift zero.

\begin{figure*}
 \hfill
 \subfigure[]{\includegraphics[width=8cm]{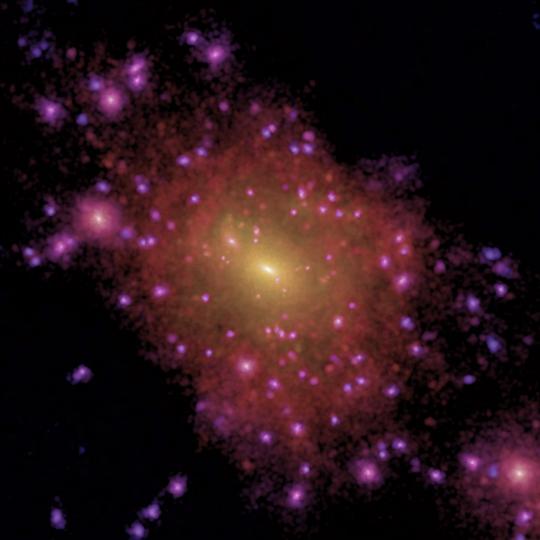}}
 \hfill
 \subfigure[]{\includegraphics[width=8cm]{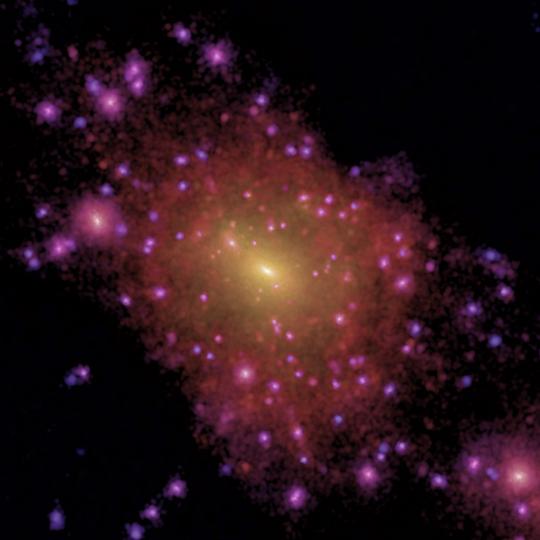}}
 \hfill
\caption{(a) Image of the reference halo at redshift zero. The projection
is in the $x_1-x_2$ plane. The side of the image measures 1~\hmpc.
The intensity of the image is scaled by the integrated square density
of dark matter, while the hue is set by the local velocity dispersion.
See  {Springel} {et~al.}(2008) for more details about this method of
making images. (b) A resimulation of the same halo discussed in
Subsection~\ref{resim-test}. }
\label{ref_pic}
\end{figure*}

  For the reference calculation we used a $12288^3$ Fourier transform
to make a set of cosmological initial conditions.  This required 35~Tb
of RAM to run and was only possible through access to the new DiRAC~II
facility at Durham. Figure~\ref{ref_pic}(a) shows a projection of the
reference halo made using the same software as was used to render the
Aquarius haloes \citep{Springel08}.  
The location and some of the
bulk properties of the halo are both defined and given in Table~\ref{halo_props}.

\subsection{How to measure convergence in the phase information\label{phase_measure}}
  We need a quantitative measure to determine how well the phase
information has been reproduced for different sets of initial
conditions.  As the true solution is not known the best we can do is
to use the reference halo as a close approximation to the truth, and
compare the halos produced from other sets of initial conditions to
the reference halo.  This assumption cannot be fully tested but we
will make some consistency checks in the next subsection to show that
it is reasonable.

 Rather than focus on how well particular properties such as the
virial mass are reproduced we will measure how well the redshift zero
particle positions are reproduced. As all simulations start with the
same particle load we can match the particles that originate from the
same location of the particle load, for any pair of simulations, and
measure their separations as a function of time.  We expect the
distribution describing the relative positions of particles in any
pair of our simulations to diverge over time. If for the end state at
redshift zero all the particle positions match well, then we can be
confident that not only will the physical properties of the halo at
redshift zero agree very well between the two simulations, but this
agreement will apply over the entire past history.  For if this were
not the case it would require a vast coincidence to have occurred --
something that we can reasonably discount.  This requirement is
considerably more rigorous than just trying to match a few physical
attributes of a single halo as these much more likely to agree
well just by chance.

Not surprisingly the degree to which the positions agree over time
depends strongly on the set of particles selected for comparison: we
observe larger differences in the redshift zero relative positions of
samples of particles chosen to be close to the halo centre, than
further out beyond the virial radius.  For our sample we will take all
particles between 200 and 300~kpc/$h$ of the potential centre of the
redshift zero reference halo. This choice is somewhat arbitrary, and
is ultimately made on aesthetic grounds: we find our test measure,
described later, shows a greater variation between the different sets
of initial conditions than a sample chosen from within the halo
itself.  This variation makes for clearer figures.  In fact the
conclusions we draw are insensitive to the sample choice provided the
particles are taken from the high resolution region.

 We define the positions of the particles relative to the halo centre
in each simulation, so the relative positions of particles between
initial conditions are insensitive to translations of the haloes. This
means the differences in the halo positions, as defined by their
potential minima, are an independent
measure.

  The distribution of the relative particle displacements between the
end states of two simulations has a very long tail to large
separations: there are always some particles that end up on opposite
sides of the halo. The majority of particles however typically have a
much narrower spread. To avoid being strongly influenced by the tail
of the distribution, which contributes little to the total mass
density, we choose the median of the distribution of separations as
the test measure and call it \MDIFF. We have checked that taking other
percentiles such as 25\% or 75\% makes no significant difference to
the rankings of pairs of simulations.  For convenience we will measure
\MDIFF\ in units of \hkpc.

  To get some intuition as to how this measure behaves we first test
it between pairs of simulations that are extremely similar.  It is
well known that the end states of N-body simulations can be very
sensitive to extremely small changes in their initial conditions
\citep{Miller1964}.  We can see this effect using \MDIFF\ as a
measure.  Taking the reference set of initial conditions we make a new
set of initial conditions by copying them and modifying the velocities
of each particle, which are represented as single precision floating
point numbers, and randomly perturb each velocity up or down by the
smallest amount possible. We use different random sequences to produce
five sets of perturbed initial conditions.  This perturbation
introduces a fractional `error' on the velocities of about one part in
ten million.  It is inevitable at least with single precision
velocities that any initial conditions will contain errors at least of
this magnitude.

We then ran these five sets of simulations to redshift zero and looked
at the median differences in the particle positions between the 10
pairs of initial conditions. The values of \MDIFF\ in \hkpc\ are shown
below:
\begin{equation*}
\begin{matrix} 
   6.4      &      &      &      \\
   6.2      &  6.0 &      &      \\
   6.3      &  6.0 & 6.1  &      \\
   6.5      &  5.9 & 5.9  &  6.1 
 \end{matrix}
\end{equation*}
 The mean differences in \MDIFF\ between all the pairs are remarkably
consistent.  In size they are slightly greater than 4\% of
$R_{200}$. Had we taken a sample of all particles within 300~\hkpc\
instead, then the differences would be about 18~\hkpc\ which is more
than 10\% of the virial radius. These differences are largely random
rather than systematic as the physical properties of the halo in these
five different versions are extremely similar to each other and to the
reference halo.  The fractional RMS variation in the quantities
$M_{200}$, $V_{\rm disp}$ and $V_{\rm max}$ are 0.1\%, 0.15\% and
0.2\% respectively.  These uncertainties give a measure of how
accurately one can reasonably expect to determine these quantities
even with extremely high quality initial conditions and a lower
limit of what to expect for \MDIFF.

\subsection{Testing cosmological initial conditions  \label{cosmo_test}}
 Having established a benchmark for the size of \MDIFF\ for virtually
identical initial conditions we will now look at the differences in
the final halo between sets of cosmological initial conditions. We
expect the quality of the phase reconstruction to depend strongly on
the size of the Fourier grids used.  The larger the Fourier grid the
more information from \WNF\ can be used to generate the phases.  As
described in the last section, all sets of initial conditions contain
additional information from a field which is uncorrelated with \WNF\
and that is introduced to restore isotropy to the initial conditions
particularly at small scales. We will call this field the `independent
field'.  We can investigate the influence of adding this independent
field further by generating an ensemble of initial conditions that
differ only in using a different realisation of the independent field.

 \begin{table}
{
\begin{tabular}{| l | l | r | c |} 
\hline
  Index     &  Name   &   1-D FFT size   & S$_1$/S$_8$         \\
\hline
   1        & Reference        & 12288    &  8 \\
   2        & Ref-alt          & 12288    &  8 \\
   3        & Cosm-6144        &  6144    &  8 \\
   4        & Cosm-6144-alt1   &  6144    &  8 \\
   5        & Cosm-6144-alt2   &  6144    &  8 \\
   6        & Cosm-6144-alt3   &  6144    &  8 \\
   7        & Cosm-6144-alt4   &  6144    &  8 \\
   8        & Cosm-6144-alt5   &  6144    &  8 \\
   9        & Cosm-3072        &  3072    &  8 \\
  10        & Cosm-3072-alt1   &  3072    &  8 \\
  11        & Cosm-3072-alt2   &  3072    &  8 \\
  12        & Cosm-3072-alt3   &  3072    &  8 \\
  13        & Cosm-3072-alt4   &  3072    &  8 \\
  14        & Cosm-3072-alt5   &  3072    &  8 \\
  15        & Cosm-3072-S$_1$   & 6144    &  1 \\
  16        & Cosm-6144-S$_1$   & 3072    &  1 \\
  17        & Ref-fp1          & 12288    &  8 \\
  18        & Ref-fp2          & 12288    &  8 \\
  19        & Ref-fp3          & 12288    &  8 \\
  20        & Ref-fp4          & 12288    &  8 \\
  21        & Ref-fp5          & 12288    &  8 \\
\hline                                  
\end{tabular}
}\caption{Cosmological initial condition sets used in
Figure~\ref{cosmo_comp}. The index is used as a label
in that figure. See main text for more details. }
\label{cosmo_sims}
\end{table}

We can make cosmological initial conditions using the information
provided by all eight of the Legendre block coefficients, or
just the $p_{000}$ block. This allows us to compare the relative
merits of using either the S$_8$ or S$_1$ octree basis functions.
From Figures~\ref{octree_powspec} and \ref{frac_gauss} we expect
to see significant differences and dependencies on the Fourier
grid size.

  Table~\ref{cosmo_sims} shows all the cosmological initial
conditions we use in this subsection.  The top set of initial
conditions in the table is the reference calculation described earlier
on. The `Ref-alt' set is identical to the reference calculation except
that it has a different realisation of the independent field.  These
two sets of initial conditions were very expensive to set up so we
have made do with just two realisations of the independent field.  The
Cosm-6144 and Cosm-3072 sets are analogous to the reference simulation
but were set up using $6144^3$ and $3072^3$ sized Fourier grids
instead.  For each of these we generate five further sets with
different realisations of the independent field as indicated by '-alt'
postfix to the names.  In addition there are two sets Cosm-6144-S$_1$
and Cosm-3072-S$_1$ which were made up using the S$_1$ octree basis
functions with $6144^3$ and $3072^3$ Fourier grids respectively.
Finally in the table the five sets of initial conditions with a `-fp'
postfix are those introduced in the previous subsection. These differ
from the reference set by the smallest differences possible for
single precision floating point representations of the particle
velocities.

\begin{figure}
\resizebox{\hsize}{!}{
\includegraphics{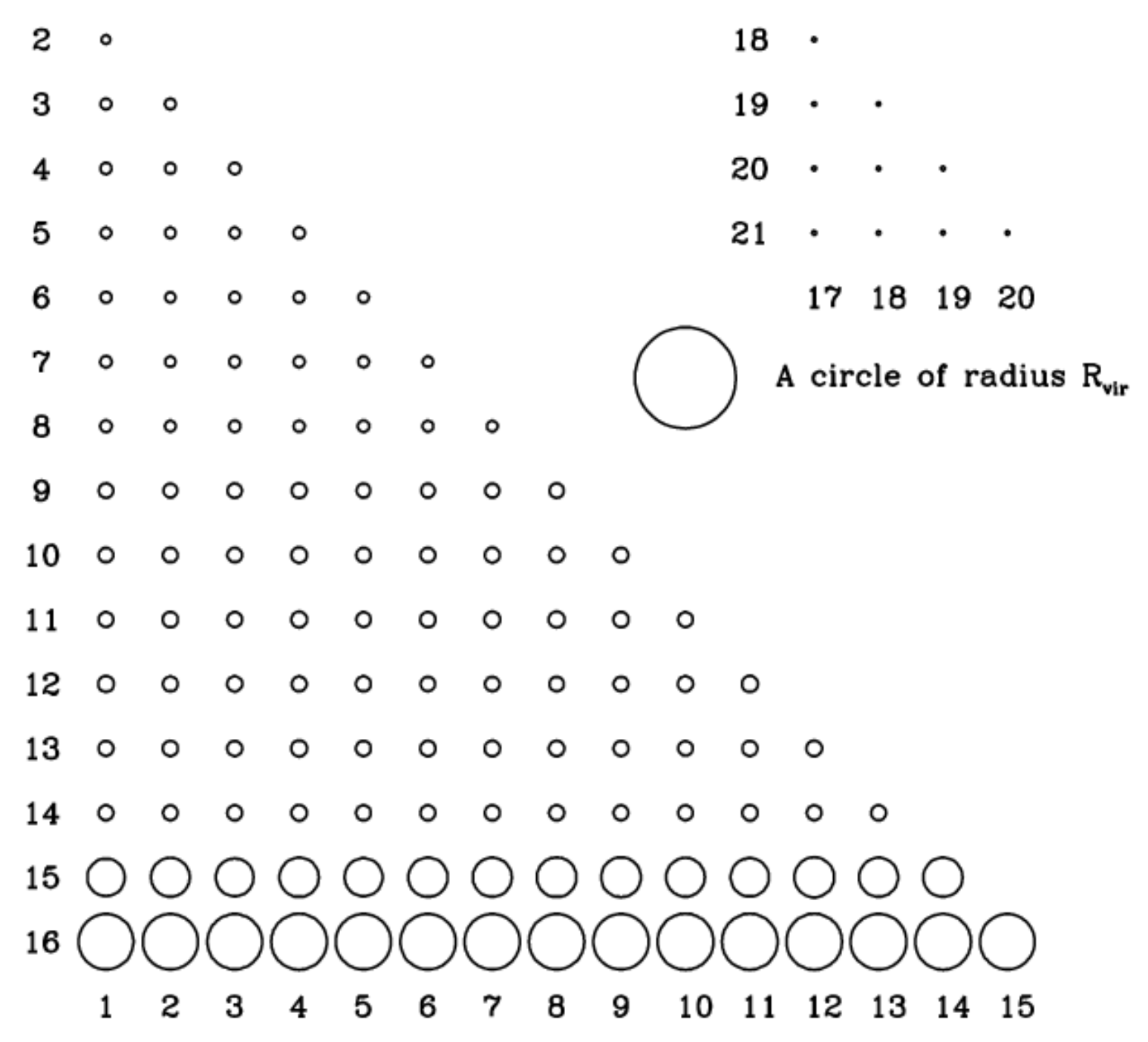}}
\caption{Each circle corresponds to a comparison made between two
completed simulations started from the initial conditions listed
in Table~\ref{cosmo_sims}. The numbers in the rows and columns
correspond to the index given in the table. The radius of each
circle is proportional to the quantity \MDIFF\ defined in 
Subsection~\ref{phase_measure}. The smaller the circle the
better the agreement. A circle of radius $R_{\rm vir}$ is
shown for scale. }
\label{cosmo_comp}
\end{figure}

  The large and small triangles made of circles in Figure~\ref{cosmo_comp} show
a set of comparisons between pairs of initial condition listed in
Table~\ref{cosmo_sims}.  The row and column numbers of each circle
correspond to the indices given in the table and show which pair of
simulations is being compared. The radius of each circle is 
proportional to the value of 
\MDIFF\ evaluated between the pair of haloes generated from the initial
conditions. For comparison the large isolated circle has a radius of
157~\hkpc, corresponding to $R_{200}$ of the reference halo.

\noindent Looking at Figure~\ref{cosmo_comp} we can observe:
\begin{itemize}
 \item  The smallest circles are those corresponding to the comparisons
between sets 17-20 and represent the smallest differences we might
reasonably expect to see between pairs of initial conditions given the
presence of single precision floating point errors.

\item The smallest circle in the main triangle is that for the pair of
simulations run from initial conditions made with the S$_8$ octree
basis functions and the largest Fourier transform size: $12288^3$, as
we would expect from the predictions of the linear power spectrum
shown in Figure~\ref{frac_gauss}.  This circle has a radius of
13.5~\hkpc\ which is about a factor of two greater than what is
potentially achievable at the floating point limit. As the separations
are in three dimensional space the associated change in volume
corresponds to a factor of ten worse.  The addition of the independent
field does have a measurable effect on the accuracy of the phases even
when using a $12288^3$ Fourier transform. However, as we will see later
that the effect on the bulk halo properties such as virial mass
and maximum circular velocity is sufficiently small as to be
indiscernible in practice.

\item Below this, the circles making rows 3-8 are comparisons between
pairs with initial conditions where one was made with a $6144^3$
Fourier grid and the other a $6144^3$ Fourier transform or
greater. These 27 circles are all very similar in size ranging from
15.9-18.0~\hkpc\ with a mean of 16.9~\hkpc. In volume terms \MDIFF$^3$
is a factor two larger when compared the results for a $12288^3$
Fourier transform.

 \item The circles in rows 9-14 consist of comparisons between pairs
with one made using a $3072^3$ Fourier grid and the other a $3072^3$
Fourier grid or larger.  Again these circles are remarkably similar in
size and distinct from those above.  They range in size from
21.5-24.9~\hkpc\ and average 23.2~\hkpc.  (\MDIFF)$^3$ is a factor 5
larger than for the pair generated with a $12288^3$ Fourier transform.

 \item Finally the bottom two rows consist of comparisons where one of
the initial condition sets were made using the S$_1$ octree basis
functions.  A $6144^3$ Fourier grid was used for the penultimate row,
and $3072^3$ grid for the bottom row. The average \MDIFF\ value for a
$6144^3$ Fourier grid is 60~\hkpc, and for $3072^3$ grid it is
86~\hkpc.  In terms of (\MDIFF)$^3$ these values are about 90 and 260
times respectively larger than using the S$_8$ octree basis functions
on a $12288^3$ Fourier transform.

\end{itemize}

 The consistency of the \MDIFF\ measure between the ensembles
of initial conditions made with $6144^3$ and  $3072^3$ Fourier
transforms suggest that the one measurement between the pair
of simulations using a  $12288^3$ Fourier transform is likely
to be representative of an ensemble, had we been able to afford
to make them. 

\begin{table*}
{
\begin{tabular}{| l | l | l | l | l | l | l |}
\hline
 Quantity       &  Reference &   Ref-fp          & $6144^3$/S$_8$     & $3072^3$/S$_8$     & $6144^3$/S$_1$  & $3072^3$/S$_1$ \\
\hline
 Number of simulations &  1    &  5       &    6        &  6    & 1   & 1\\
 $M_{200}$/ $10^{10}h^{-1}$~\msun      &  91.61    & $91.59\pm0.09$ &  $91.52\pm0.12$ & $91.41\pm0.12$ &  90.08     &   86.16    \\
$ V_{\rm disp}$/ km~s$^{-1}$ &  102.72    & $102.89 \pm0.17$  &  $102.80 \pm0.27$  & $103.26 \pm0.18$   &  102.28    &   99.72   \\
 $ V_{\rm max}$/ km~s$^{-1}$  &  181.71   & $181.83 \pm0.30$  &  $181.70 \pm0.37$  & $182.23 \pm0.40$   &  182.22    &   179.27   \\
\hline                                  
\end{tabular}
}
\caption{Physical parameters of the dark matter halo at redshift zero for
a series of simulations running using initial conditions listed in
Table~\ref{cosmo_sims}. The column headings refer to initial
conditions with indices in the table as follows: Reference 1; Ref-fp
17-21; $6144^3/{\rm S}_8$ 3-8; $3072^3/{\rm S}_8$ 9-14; $6144^3/{\rm
S}_1$ 16; $3072^3/{\rm S}_1$ 15.  In columns where there is more than
one simulation the numbers given are the average of the quantity for
the sample together with the estimated RMS about the average.}
\label{cosmo_phys_comp}
\end{table*}

 The trends seen in Figure~\ref{cosmo_comp} are consistent with what
would be expected: using a larger Fourier grid leads to a better match
to the reference simulation; and using the S$_8$ octree basis
functions is much better than using the S$_1$ set.  The \MDIFF\
measure shows these trends very clearly.  There is no overlap in the
sizes of the circles between the sets of rows described above. While
the trends in \MDIFF\ have a clear physical interpretation, it is not
obvious how these values are related to errors in reproducing simple
measured bulk properties for a halo.
 
 In Table~\ref{cosmo_phys_comp} we compare the values of $M_{200}$,
$V_{\rm disp}$ and $V_{\rm max}$ for the redshift zero halo from all of
the runs listed in Table~\ref{cosmo_sims}.  Where there are several
similar initial conditions we compute a mean and RMS about that mean.
With only quite small samples the estimates of the RMS themselves have
a significant error.

  We can see that the cosmological initial conditions made using S$_8$
and a $6144^3$ Fourier transform have properties that are close to the
reference calculation. The estimated RMS values are comparable or
smaller than the differences between the means.  For the results using
the $3072^3$ Fourier transform the level of agreement is still very
good, but the differences between the mean and the reference
calculation are larger. 

  This situation for the two halo simulations run from initial
conditions made using S$_1$ octree basis functions is much less
satisfactory. The two haloes produced from these initial conditions
differ from the reference halo by a much wider margin then any other
halo.  As expected using a larger Fourier grid does improve the
quality of the initial conditions.  The result for a $6144^3$ Fourier
transform does match $V_{\rm max}$ well, but this likely a coincidence
as we know the underlying particle distributions are significantly
different.

The \MDIFF\ statistic is by construction insensitive to whether the
halo potential centres agree of not.  Comparing the absolute positions
of the halo centres we find that all simulations run from initial
conditions created using S$_8$ agree in the position of the potential
minimum to better than 2~\hkpc, while the two simulations run
from initial conditions made using S$_1$ differ from each other
and all other simulations by more than 10~\hkpc.

  The degree of agreement in the physical properties between the
reference halo and the others is broadly consistent with the trends in
\MDIFF\ although less clear cut.  The cosmological initial
conditions using the S$_8$ octree basis functions with a $6144^3$
Fourier grid appear to be indistinguishable from the reference halo
for the bulk halo properties $M_{200}$, $V_{\rm disp}$ and $V_{\rm
max}$.  The agreement with those made with $3072^3$ grid is extremely
close and at the sub-percent level.  Using S$_1$ octree basis
functions, however, leads to much poorer results with properties such as
$M_{200}$ differing by more than 4\% when a $3072^3$ grid is used. 

We conclude from these comparisons of the bulk properties of the halo
that it is possible to generate high quality cosmological initial
conditions that accurately reconstruct the phase information given by
\WNF.  The results obtained using the S$_8$ set of Legendre blocks are
sufficiently good that it is not obvious that going to more
complicated octree basis function expansions such as S$_{27}$ would
reproduce the bulk properties of the halo any more accurately.  It is
clear however that just using the S$_1$ octree basis functions gives
poor quality initial conditions and is therefore not recommended.

\subsection{Testing the resimulation method \label{resim-test}}

  Having established the quality of the initial conditions made using
the cosmological method we can now use them to test the quality of the
resimulation initial conditions made using the method outlined in
Subsection~\ref{resim_method}. The main goal of this subsection,
and indeed the section, is to demonstrate that it is possible
to make good quality resimulation initial conditions with \WNFIAN\ phases.
To do this we will simply compare the bulk properties of a 
resimulated version of a halo to the reference halo and show they
agree at the sub-percent level.  We will then use the more 
rigorous \MDIFF\ measure to check this conclusion.

 There are quite a few choices that need to be made when setting up
resimulation initial conditions. In this subsection we detail these
choices without justification. We explore in the next subsection how
varying these choices affects the quality of the initial conditions.

 For these resimulation initial conditions we use 4 Fourier meshes
centred around the high resolution region with linear sizes of 1, 1/2, 1/4
and 1/8 of the periodic volume. We use a $768^3$ Fourier transform for
all meshes.  The memory requirement of the \ICgen\ code for this grid
size is about 10 Gbytes which means the code can comfortably fit on a
modern supercomputer node.  We will take the value of the `X'
parameter, introduced in Subsection~\ref{resim_method} to be 4.  We
will use S$_8$ octree basis functions.

  Figure~\ref{ref_pic}(b) is an image of the resimulated
halo. Clearly, it resembles the reference halo
closely. Table~\ref{resim_example_props} gives some of the bulk
properties of this halo and shows how it differs from the reference halo.
The differences are remarkably small - at the subpercent level. We
find similar levels of agreement for resimulations of other haloes
using the same methods so these numbers can be taken as typical. 
We conclude that is is possible to make high quality resimulation
initial conditions from \WNF.

\begin{table}
{
\begin{tabular}{| l | c | c |} 
\hline
  Property            &   Value(s)                &   $\Delta$Reference      \\
\hline
  Centre              & (43.1712, 49.9135,2.2093) & (0.0020\ ,0.0035\ ,0.0023) \\  
  $M_{200}$           &  91.32                    & -0.29       \\
  $M_{\rm vir}$       &  109.96                   & -0.06       \\
  $R_{200}$           &  157.77                   & -0.18       \\
  $R_{\rm vir}$       &  213.43                   &  0.04       \\
  $V_{\rm disp}$      &  102.89                   &  0.17       \\
  $V_{\rm max}$       &  181.70                   & -0.01       \\
    $\lambda^\prime$  &  0.0133                   & 0.0000      \\
  $(c1,c2,c3) $       &  (-0.216,-0.641,-0.736)   &  $0.75^{\circ}$  \\
\hline                                  
\end{tabular}
}\caption{ The properties of the resimulated halo and a comparison to
the reference halo. The quantities and their units are explained in
Table~\ref{halo_props}. The third column gives the difference between
the resimulated halo and the reference halo. For the directional
cosines the angle between the spin directions is given.}
\label{resim_example_props}
\end{table}

 This conclusion is supported by the measured value of \MDIFF\ between
the reference halo and the resimulated halo.  The value is 22.8~\hkpc,
which is slightly better than the difference in the mean between the
reference halo and the haloes generated from cosmological initial
conditions made with a $3072^3$ Fourier transform.  By this measure
the resimulation initial conditions are of similar quality as can be
created using single very large Fourier transforms. In effect we can
say that the resimulation method is able to produce initial conditions
of comparable quality to what can be achieved using the Fourier method
developed in the 1980s to model cosmological volumes, but at
considerably lower computational cost.

  Encouraging as these results are we can see from the value of
\MDIFF\ that it may be possible to make better initial conditions than
we have achieved. Compared to the difference in \MDIFF\ between the
two sets of cosmological initial conditions made with a $12288^3$
Fourier transform the difference in (\MDIFF)$^3$ is a factor of 4.  This
is a reasonable comparison as the mesh spacing of the inner most grid
of the resimulation initial conditions is the same as for the
cosmological initial conditions made with a $12288^3$ Fourier
transform so the same information from \WNF\ is used for the inner
region. 

\subsection{Sensitivity to parameters \label{resim_param_search}}

 \begin{figure}
\resizebox{\hsize}{!}{
\includegraphics{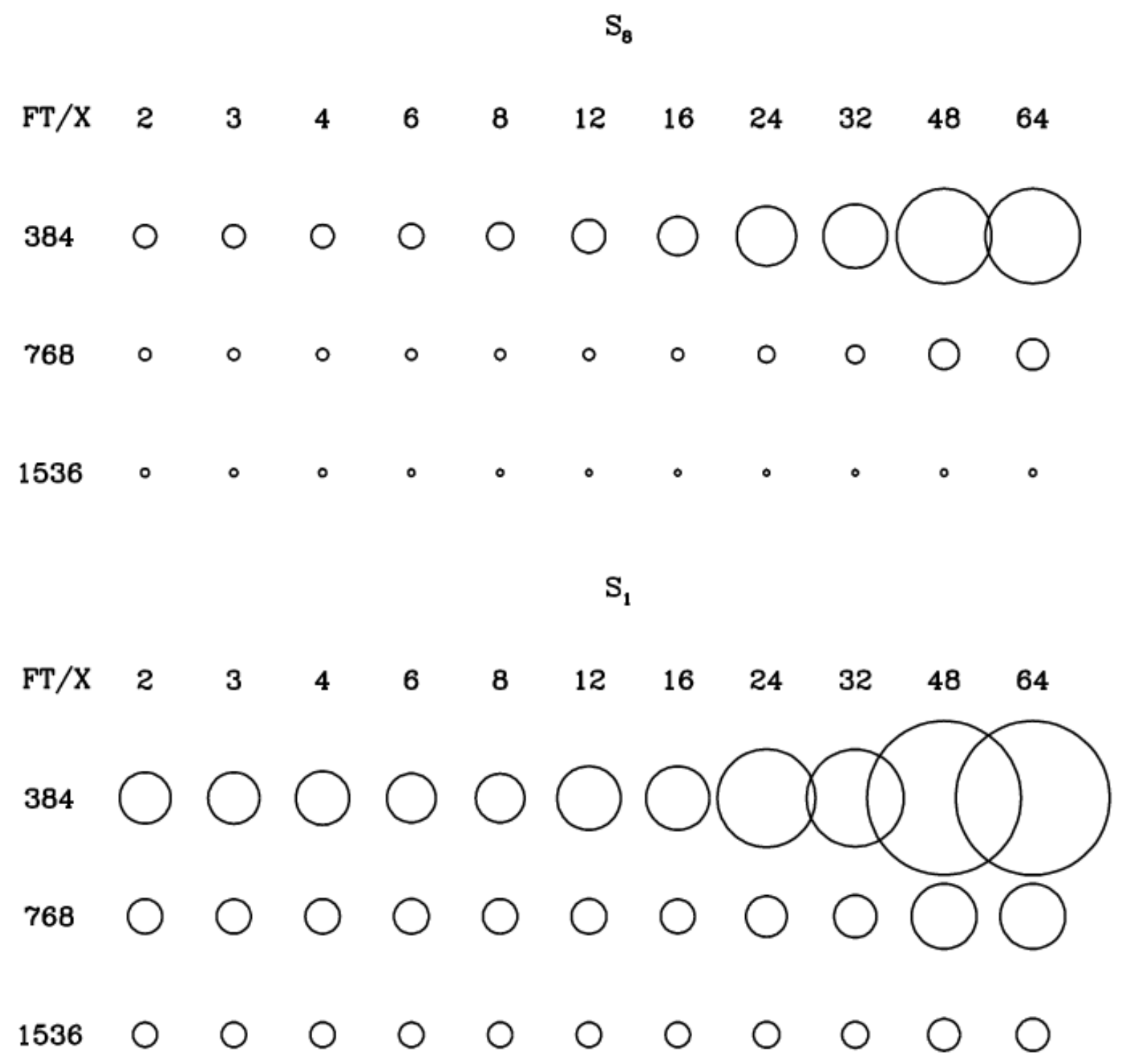}}
\caption{Each circle corresponds to a comparison between the
reference halo and a halo produced by running a set of resimulation
initial conditions, set up using the parameter values shown labelling
the rows and columns. The rows are labelled by the one dimensional
size of the Fourier transform used, the columns by the value of the X
parameter.  The upper set of circles use the S$_8$ octree basis
functions, the lower set the S$_1$ ones. The smallest circle
corresponding to FT=1536, X=12, S$_8$ has a radius of 20.0~\hkpc.
By contrast the largest circle has a radius of 260~\hkpc.}
\label{param_scan}
\end{figure}

  In this subsection we show how the choice of the Fourier mesh size,
the `X' parameter (defined by Rule 4 of
Subsection~\ref{resim_method}), and the choice of S$_8$ or S$_1$
octree basis functions affect the accuracy with which the phase
information can be reconstructed in resimulation initial conditions.
We expect the results to improve with the size of the Fourier grid.
This is both because more \WNF\ information is used and because
each given octree basis function is sampled more finely by the Fourier
grid. We can maximise the sampling of the octree basis functions by
applying rule 3 of Subsection~\ref{resim_method}), that states we
should place octree functions on the finest possible grid.  However, we
introduced a fourth rule, which forbids large octree basis functions
being placed close to the boundaries of any but the outermost grid.
The reasoning for this was that placing large octree basis functions
near the boundaries of the inner grids would lead to larger edge
effects. A parameter X was introduced so that the perpendicular
distance of any but the smallest octree basis functions should not be
placed within a perpendicular distance to the boundary that is X times
its own size.  We expect for very large values of X this rule will
tend to place octree basis functions on coarser grids, and will lead
to poorer reconstruction of the phases because of poor sampling of the
octree functions. For very small values of X boundary effects may also
have a negative effect.  Finally, we expect better results using the
S$_8$ octree basis functions compared to the S$_1$ set as we saw for
cosmological initial conditions.

 We have made sets of initial conditions with a wide range of 
combinations of Fourier grid size, X and type of octree basis
function: the Fourier grid sizes used are $384^3$, $768^3$, $1536^3$;
the values of X are 2,\;3,\;4,\;6,\;8,\;12,\;16,\;24,\;32,\;48 and 64; and with S$_8$
and S$_1$ octree basis functions.  We then run each set of initial
conditions to redshift zero and have computed \MDIFF\ with respect to
the reference halo.

 The results of those comparisons are shown in
Figure~\ref{param_scan}. As expected we see a clear trend with Fourier
grid size, and between using the S$_8$ and S$_1$ basis functions.  The
best result using S$_1$ and a $1536^3$ Fourier transform is still
poorer than the best using S$_8$ and a $384^3$ grid. The former
initial conditions are also almost two order of magnitude more
expensive to generate. This confirms for resimulation initial
conditions the conclusion we had already reached for cosmological
simulations that the $S_1$ basis functions are unsuitable for accurate
work.

 Concentrating on the S$_8$ results, we see in Figure~\ref{param_scan}
that larger values of X are clearly disfavoured for the $384^3$,
$768^3$ grids. There is almost no trend for a $1536^3$ grid. For this
grid size the value of \MDIFF\ does show an increase for X=128 and 256
(not shown) where that values of \MDIFF\ of 26.9~\hkpc\ and
37.3~\hkpc\ respectively. Choosing a maximum value of X that depends
on the grid size give an upper limit to the X parameter. Using a
value ${\rm X}<M/24$, where $M$ is the Fourier grid choice appears
a safe choice, with larger values clearly disfavoured.

  While large values of the X parameter give poor results, the range
of X explored does not show any convincing evidence for boundary
effects being important at low values of X.  It is not true however
that the lowest values of \MDIFF\ occur for the lowest values of X, it
is just that there is no clearly distinct minimum in \MDIFF\ values.
We conclude from these tests that the quality of the resimulation
initial conditions are  not that sensitive to precisely which grid the
octree basis functions are placed, above a mimimum sampling.  The
choice of $4\le{\rm X}\le12$ appears to work well for all grid sizes
we have tested.

\begin{table*}
{
\begin{tabular}{| l | l | l |}
\hline
  Simulation  &     Reference     &     Phase Descriptor \\
\hline
  \DOVE\     &   Not yet published         & [Panph1,L16,(31250,23438,39063),S12,CH1292987594,DOVE]\\
  \Millgas\  &   \cite{Guo_13}    & [Panph1,L11,(200,400,800),S3,CH439266778,MW7]\\
  \MXXL\     &   \cite{Angulo2012} &  [Panph1,L10,(800,224,576),S9,CH1564365824,MXXL]\\
 \hline                                  
\end{tabular}
}
\caption{The phases of several recent Virgo Consortium simulations. The
text phase descriptor gives the location of the phase information in \WNF.
See Section~\ref{PUB-CONV} for details. The \Millgas\ simulation has
been added to the Millennium database (Guo {et~al}, 2006).}
\label{wn_publish}
\end{table*}

The smallest circle in the diagram corresponds to X=12 and a $1536^3$
Fourier transform and has a value of \MDIFF\ is 20.0~kpc/$h$. This
value is intermediate between that found for the $3072^3$ and $6144^3$
cosmological initial conditions.  The run time to make the
resimulation initial conditions with a $1536^3$ Fourier grid is
however longer than the N-body simulation takes to go from redshift 63
to redshift zero. This seems rather excessive for a real
application. For this reason we chose to use a $768^3$ Fourier
transform for the last subsection to demonstrate the quality of the
resimulation initial conditions. With this grid size the lowest value
of \MDIFF\ is 22.8~\hkpc\ for X=4. These particular {\sc 2lpt} initial conditions
take about twenty minutes to create on a single 16 core node.  Making
Zeldovich initial conditions would be considerably quicker.  The
N-body simulation on the same node takes about two hours.  For more
sophisticated simulations of structure formation modelling
hydrodynamics and complex subgrid models the run time may be much
longer than for the pure N-body simulation. In such cases the
fractional costs of making the initial conditions become relatively
small.

\section{Publishing phase information\label{PUB-PHASE}}
 All that is needed to specify the phase information of 
initial conditions generated from \WNF\ is the spatial
location of the phases within the volume.  To use this
information to resimulate a region of interest within a
simulation requires specifying the position and dimensions
of the region at high redshift.

 In Subection~\ref{PUB-CONV} we define a convention for specifying
phases taken from \WNF\ and using this convention publish the phases
of three cosmological volumes run by the Virgo Consortium.  In
Subsection~\ref{PUB-SIMS} we give the locations and sizes of a region
within each of these volumes out of which forms a dark matter halo at
redshift zero.  We show images and give the positions and a few
properties of these haloes for reference.

\subsection{How to publish Panphasia phases\label{PUB-CONV}}

 For the purposes of designing a convention for publishing phase
information we will assume the phase information for all simulations
is defined by specifying either a cube or a cuboid within \WNF\ made
of complete octree cells.  The location of a cube within an octree
requires five integers: one integer to define the smallest level in
the octree that is possible; three integers to specify the location of
the corner cell closest to the origin, using the Cartesian
coordinates described in Section~\ref{octree} to label the cells; and
one integer to give the side length of the cube in units of the octree
cell at the given level.  For a general cuboid, or one with two
different side lengths, we will require three side lengths to
be given.

  Taking these integers we define two text phase descriptors incorporating
these numbers: one for cubic regions and one for a general cuboid.
Because making an error in the value of any of the integers in a
descriptor would change the phase information, we include an
additional check number in the descriptor.  To be useful this check
integer must depend on all of the integers that define the phase.
While having an error check can avoid some human errors it is no safeguard
against simply using the wrong descriptor. It is desirable for the
descriptor to also include a human readable name that can be readily
associated with a particular simulation volume.

The check number we have  selected combines the random
number states associated with the three corner cells adjacent to the
corner cell nearest the origin and the name of the phase descriptor.
The full details are given in Appendix~\ref{MAP-OCT}.

 For a cubic region we define a plain text phase descriptor:
\begin{equation*}
 [{\rm Panph1,L}\#_0,(\#_1,\#_2,\#_3),{\rm S}\#_4,{\rm CH}\#_5,{\sc string}],
\end{equation*}
where each symbol $\#$ represents an integer: $\#_0$ is the octree
level, $(\#_1,\#_2,\#_3)$ are the Cartesian coordinates of the corner
cell nearest the origin, $\#_4$ is the side-length and $\#_5$ is the
check number.  For a cuboidal region we define a second phase descriptor:
\begin{equation*}
 [{\rm Panph1,L}\#_0,(\#_1,\#_2,\#_3),{\rm D}(\#_4,\#_5,\#_6),{\rm CH}\#_7,{\sc string}],
\end{equation*} where the three side lengths are given by $(\#_4,\#_5,\#_6)$
for each Cartesian direction, and $\#_7$ is again the check number.
Finally {\sc string} is a text string, again without spaces, naming the
particular realisation of the phases. This should be distinctive as
this is the best protection against using the wrong descriptor by
accident.
 
  There are no spaces within the phase descriptor and the type of brackets
punctuation, and cases of the letters should be observed. The string
`Panph1' is intended to help identify the descriptor and to make it
possible to make a text search for \WNF\ descriptors.  The `1' allows
for the possibility of extending or adapting the format in the future.

 A code to randomly generate phase descriptors, including the check digit,
 is included with the public release of \WNF.  

  Table~\ref{wn_publish} gives the phase descriptors for three cosmological
simulations run by the Virgo Consortium including the \DOVE\
simulation from which we resimulated the reference halo. In the next subsection
we give examples of a halo that can be resimulated from each of these
volumes.

\subsection{Examples haloes for resimulation.\label{PUB-SIMS}}

\begin{figure}
\resizebox{\hsize}{!}{
\includegraphics{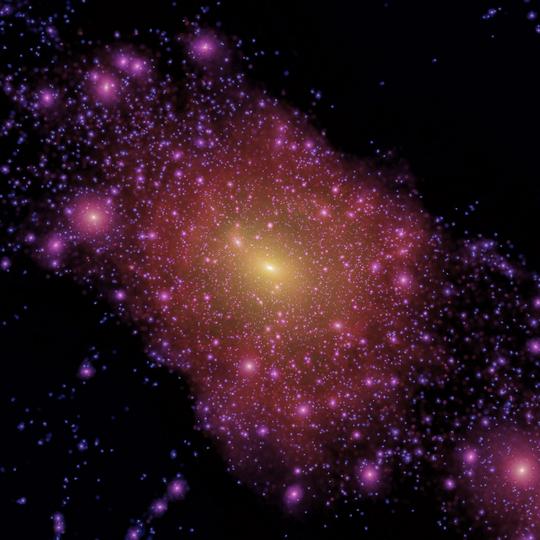}}
\caption{High resolution version of \DOVE\ reference halo
used in this paper. The projection is the same
as in Figure~\ref{ref_pic}.
}
\label{resim_ref_pic}
\end{figure}

\begin{figure}
\resizebox{\hsize}{!}{
\includegraphics{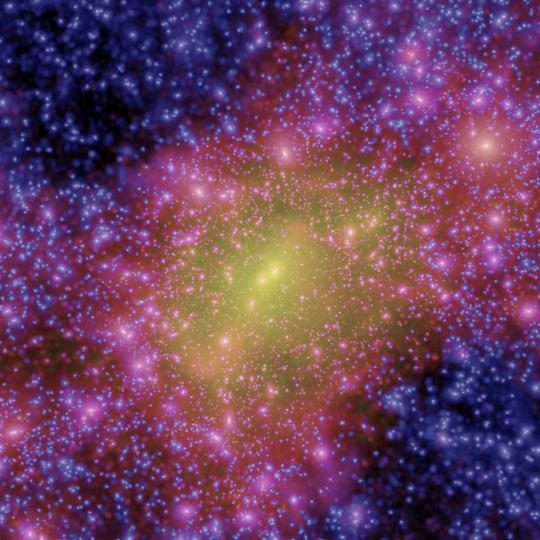}}
\caption{High resolution version of the \Millgas\ reference halo at
redshift zero.  The projection is in the $x_1-x_2$ plane and the side
length of the image is 10~\hmpc. This is the most massive cluster in
the \Millgas\ volume at redshift zero.  }
\label{resim_millgas_pic}
\end{figure}

\begin{figure}
\resizebox{\hsize}{!}{
\includegraphics{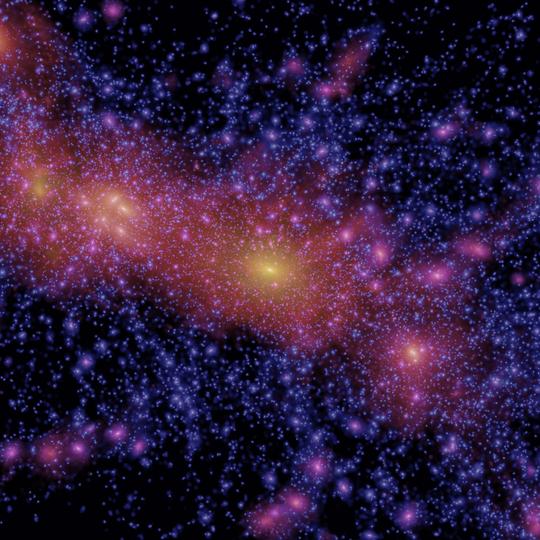}}
\caption{ High resolution resimulation of a reference halo in the
\MXXL\ simulation. The projection at redshift zero is in the $x_1-x_2$
plane, and the image measures 15\hmpc on a side. }
\label{resim_mxxl_pic}
\end{figure}

 For anyone wanting to implement \WNFIAN\ phases in a new code it is
desirable to have some test cases to check that the code is working
correctly.  In Table~\ref{wn_resim_examples} we give the locations and
sizes of a single sphere within each of the \DOVE, \Millgas\ and
\MXXL\ volumes at high redshift.  A resimulation of these spheres
results at redshift zero in the formation of a prominent dark matter
halo selected from these cosmological volumes. Images of these haloes
at redshift zero, projected onto the $x_1-x_2$ plane are shown in
Figures~\ref{resim_ref_pic}, \ref{resim_millgas_pic} and
\ref{resim_mxxl_pic}.  The \DOVE\ and \Millgas\ volumes have the same
cosmological parameters, given in Table~\ref{cosmo_params}.  The
\MXXL\ parameters are different and are given in \cite{Angulo2012}. We
use the same coordinate system to describe these locations as for
\WNF: the coordinates are non-negative and the origin marks one corner
of the volume.

 The redshift zero properties of these haloes, as determined by a
resimulation using the methods described in this paper, are given in
Table~\ref{wn_resim_properties}.  The halo in the \DOVE\ volume is a
resimulation of the same volume as was used for the reference halo in
Section~\ref{SIM-TESTS}, but with a factor of about thirty more
particles.  The properties of this halo are very similar to those
obtained in the reference calculation. As this resimulation has
additional small scale power it does not follow that the value of
$M_{200}$ should be precisely reproduced.  The difference we observe
between the two versions of the halo at different resolutions is
similar in size to that seen in \cite{Springel08} between the Aq-A-5
and Aq-A-4 resimulations.  This particular halo has also been
simulated at very similar numerical resolution to the example given in
this subsection with initial conditions made using a single $6144^3$
Fourier grid. The properties of the two higher resolution versions of
this halo are a very good match to each other with smaller difference
than seen between versions of the halo simulated with very different
particle numbers.

 The haloes in the \Millgas\ and \MXXL\ volumes are both in the
cluster mass range. Both these haloes are in the process of formation
and are far from equilibrium.  Each is resolved with about 8 million
particles within $R_{200}$.

\begin{table}
{
\begin{tabular}{| l | l |}
\hline
  Simulation  &   Position and size of  halo Lagrangian region. \\
              &    \hmpc \\
\hline
  \DOVE\     &   [(41.11,48.80,3.00),4.5] \\
  \Millgas\  &   [(307.46, 52.51,434.33),46]\\
  \MXXL\     &   [(1806.8,1207.5,1617.9),35]\\ 
 \hline                                  
\end{tabular}
}
\caption{Locations of a sphere within each volume from which a sizable
dark matter halo forms.  The first three digits within the round
brackets mark the centre of the sphere, while the fourth gives the
radius of the sphere. The coordinates of the simulation volume include
the origin and are non-negative.}
\label{wn_resim_examples}
\end{table}

\begin{table}
{
\begin{tabular}{| l | l | l | l | l |}
\hline
 Halo Property   &   Unit  &  \DOVE\ & \Millgas\ & \MXXL\ \\
\hline
  $x_1$         &  \hmpc           &    43.174  &  303.06  &      1802.15 \\
  $x_2$         &  \hmpc           &    49.918  &  52.90   &      1205.52 \\
  $x_3$         &  \hmpc           &    2.203  &  423.55  &      1614.56 \\
  $m_p$         &   $10^{6}$\hmsun  &   0.115   &  133.54   &      51.65 \\
  $M_{200}$     & $10^{10}$\hmsun   &    90.81   &  117756  &      42433 \\
  $R_{200}$     &   \hkpc          &    157.47  &    1717 &        1222 \\
  $V_{\rm disp}$& km~s$^{-1}$       &      104.8 &    1054  &       737 \\
  $V_{\rm max}$ & km~s$^{-1}$       &      182.1 &    1673  &       1305 \\
 \hline                                  
\end{tabular}
}
\caption{Location and a few properties of reference haloes at redshift
zero. The position $(x_1,x_2,x_3)$ is the location of the particle
with the lowest potential as determined by \SUBFIND. The
quantity $m_p$ is the mass of the particles in the high resolution 
region of the resimulation. The other quantities are defined
in the caption of Table~\ref{halo_props} }
\label{wn_resim_properties}
\end{table}

\section{Overview of the public code to generate \WNF\label{CODE-DES}}
 In this section we give a brief overview of the code to generate
\WNF. The full details are given in the companion paper \companion\
which being on the arXiv may be updated after this paper is published.  
The following therefore provides only a very general overview.

From the point of view of adding \WNF\ phases to an existing code for
making initial conditions there are just two subroutines that need to
be called directly. The first is an initialisation routine which takes
the \WNF\ phase descriptor.  The second is an evaluation routine that
returns values of the field itself for a location chosen by the user.

 The routines are serial codes. They take very little memory however
so there is no significant cost to a parallel code in having multiple
instances of the code -- with one belonging to each mpi process for
example. The code is therefore easy to use in parallel applications:
different parts of the white noise field can be computed completely
independently. The main parallel programming needed to incorporate
these routines is to ensure that the white noise field is assigned to
the correct locations as dictated by the way the relevant grids are
distributed in parallel by the application code.  As discussed in
\companion\ the speed of the code does depend significantly on the
precise ordering of successive accesses to octree cells and it is
therefore important to consider the effects of the access pattern when
adding \WNF\ to a parallel code.

 For making cosmological conditions these two subroutines are
especially simple. The initialisation routine needs the \WNFIAN\
phase descriptor, and the size of the grid that will be used to make the
initial conditions.  The latter is needed to decide which level of the
octree to sample the region of \WNF\ selected by the descriptor.  The
evaluation routine just needs to be called with three integer
Cartesian grid coordinates, and returns nine Gaussian pseudorandom
numbers - all drawn from a distribution with unit mean and unit
variance.  The first eight of these are proportional to the expansion
coefficients of \WNF\ expanded in the eight Legendre block functions,
while the ninth independent value can be used to construct a field
that is independent of \WNF.

 For resimulation initial conditions a more general initialisation
routine is provided which requires a refinement within the
cosmological volume to be specified.  The evaluation routine in this
case returns the same nine values, but now as a function of three
integer coordinates that are defined relative to the refinement
origin. This function operates in the same way even if the refinement
is wrapped by periodic boundary conditions across a simulation
coordinate boundary.  In fact the halo we resimulated from the \DOVE\
volume is located close to a coordinate boundary so this feature is
tested in this paper. 

The refinement evaluation routine also takes values for a minimum and
maximum octree level. The values of the eight Legendre blocks returned
are calculated using only the octree functions over the range of
octree levels specified. In this way the user can decide how to place
the octree basis functions. The method for placing the octree basis
functions used by \ICgen\ for setting up the resimulations in this
paper, is determined by a subroutine which gives suggested minimum and
maximum values for the octree levels as a function of position in the
grid. This subroutine is included with the public code.

 As well as the routines to evaluate \WNF, we also provide a routine
to choose a random region from \WNF\ and generate a phase descriptor. For
this purpose it is assumed that \WNF\ itself has a physical size.  The
user must specify both the physical size of the cosmological volume,
and the required dimension of volume (assumed to be a cube) measured
in grid cells.  The code uses the user supplied information together
with the unix timestamp to generate a descriptor.

 It is assumed that the root cell of \WNF\ measures 25000~\hgpc\ on a
side. This gives a volume which is about ten billion times the current
Hubble volume for \LCDM.  At the same time the mass associated with
the smallest defined octree cells at level 50 of the octree has a
corresponding mass of about $10^{-12}$\hmsun which is below the
cut-off scale for WIMP dark matter candidate which is estimated to be
around $10^{-7}$\hmsun \citep{Hofmann_01}. 

 Finally as an example we add \WNF\ to a serial public initial conditions
code described in \cite{Crocce_06}\footnote{We thank Martin Crocce for
permission to include this modified code with our software.}

\section{Summary and discussion.\label{SUM-PAP}}
  In this paper we describe a new way for setting the phase
information for Gaussian initial conditions for cosmological
simulations and resimulations of structure formation. This work builds
upon the idea of using a real-space white noise field to define the
phase information -- a method put forward by \cite{Salmon96} and
implemented by a number of authors including
\cite{Pen97,Bertschinger01,Hahn_11}. We have developed a way of
defining Gaussian white noise fields in terms of a basis function
expansion using purpose designed orthogonal basis function sets that
have a hierarchical structure based around an octree. Vast
realisations of Gaussian white noise fields can be easily created by
assigning expansion coefficients, create by a pseudorandom number
generator, systematically to the space of basis functions. Using a
pseudorandom number generator that allows rapid access to the any part
of the sequence results in the creation of what is effectively an
objective Gaussian white noise field sampled over a very wide range of
spatial scales, any part of which is readily accessible.

 We have chosen a particular set of octree basis functions that we
find on the basis of tests to be most suitable for making cosmological
initial conditions.  This choice is a compromise between the accuracy
with which the phases of large-scale modes can be reproduced from a
finite number of octree functions, and the computational cost of
evaluating the white noise field. We have found that the simplest
choice with seven distinct functional forms for the octree basis
functions is unsuitable for accurate simulation work. We show through
resimulation tests that a choice based on fifty-six functional forms
(that can however be expressed in terms of eight more primitive
functions) is sufficiently accurate as judged against the 
published results of state of the art resimulations of
dark matter haloes.

 We have created a particular realisation of a Gaussian white noise
field called \WNF. This uses our preferred octree basis functions, with
expansion coefficients derived from a commonly used pseudorandom
number generator that passes very strong random number tests.  We use
almost the entire period of the generator to create a realisation with
fifty octree levels which is able to define phase information over
fifteen orders of magnitude in linear scale.  We make \WNF\ public by
publishing in a companion paper, \companion\ a code to compute \WNF.
Small sub-regions of this larger field are suitable for setting the
phases for cosmological simulations.  The phases for these simulations
can themselves be published by pointing to the location in \WNF, from
which the phase information was taken.

 To help with this we have defined a convention for publishing phase
information for cosmological simulations set up using \WNF.  We have
published the phases of three cosmological volumes run by the Virgo
Consortium.  As a guide for anyone wanting to add \WNFIAN\ phases to
an initial conditions code we have given the the locations and
properties of three dark matter haloes that can be resimulated within
each of these cosmological volumes.

 In order to demonstrate that it is possible to create high quality
resimulation initial conditions using phases from \WNF, we have
developed a method using Fourier transforms to do the numerical
convolutions of the white noise field required to make initial
conditions with a given cold dark matter linear power spectrum. We are
able to show these methods work well by essentially making the same
initial conditions using two different methods: as cosmological
initial conditions using very large Fourier transforms -- essentially
the Fourier method in use since the 1980s, and by the new resimulation
method described here.  We find by looking at the properties of a dark
matter halo at redshift zero that it is possible to recover the final
positions of the particles to similar accuracy when using the
resimulation method or the Fourier method.  Similarly a number of halo
properties are reproduced consistently between the two methods to
sub-percent accuracies. At the same time these tests do show there is
some room for improvement in the resimulation initial conditions.  It
may well be that the very accurate multi-grid methods developed by
\cite{Hahn_11} for the {\sc music} code can be applied successfully to
\WNF\ and yield more accurate resimulation initial conditions that the
methods described in this paper.  This paper provides a guide on how
to test the quality of resimulation initial conditions made using
\WNFIAN\ phases. Any new implementation can be tested on the reference
halo studied in Section~\ref{SIM-TESTS} and so can be compared
directly with the implementation applied in this paper to the \ICgen\
code.

 While it is important that it is possible to make accurate
resimulation initial conditions from \WNF, this is not necessarily its
most important feature. The fact that using \WNF\ allows the phase
information to be published by giving a short phase descriptor has potential
benefits for all those involved in simulations of cosmological
structure formation from Gaussian initial conditions. Using \WNF\
provides a convenient way to keep track of how simulations were set
up, and makes it possible for others to reproduce and check published
simulation results, and to exploit existing simulations. It should
also make it easier to apply very different numerical techniques to
standard problems, for example in the field of galaxy formation, by
using \WNF\ as a convenient way to define the initial conditions.

 These wider benefits will only accrue if \WNF\ is commonly used. This
requires two developments.  Firstly \WNF\ would need to be used when
setting up large cosmological volumes - particularly where these
simulations or products derived from them are made publically
available.  Currently there are no simulations using \WNFIAN\ phases
on the MultiDark database \citep{Riebe_2011}, and just one on the
Millennium database \citep{Lemson_2006}.  However the Virgo consortium
is now using \WNF, so this situation will improve in the future.
Secondly \WNF\ needs to be added to existing initial conditions codes.
It is relatively easy to add to codes that make cosmological initial
conditions, but it will require some effort from a relatively small
set of people to make \WNF\ phases available in all existing
resimulation codes.

 If both these developments can be achieved, then anyone could use
these databases to select samples of objects for resimulation.  An
alternative to this however, also requiring investment of effort,
would be for those providing the databases to provide a service that
serves resimulation initial conditions in the appropriate format to
the users.

 Finally while this paper and its companion, \companion, are intended
to act as a self-contained guide on how to add \WNF\ to other initial
conditions codes, the author would be more than happy to provide help
and advice to anyone interested in adding \WNFIAN\
phases to their codes.

\section*{ACKNOWLEDGEMENTS} I would particularly like to thank Stephen
Booth of the Edinburgh parallel computing centre for advice on
pseudorandom number generators and kindly giving me his implementation
of the \MRG\ pseudorandom number generator. Thanks to BJL, CSF, SMC,
MDS, SDMW and VRS for comments on drafts or parts of drafts of this
paper.  Thanks also to Mark Lovell who made the striking colour
images.  This work was supported by the STFC rolling grant
ST/F002289/1 and made use of the DiRAC I and II facilities at Durham,
that are jointly funded by STFC, the Large Facilities Capital Fund of
BIS and Durham University.

\bibliography{wn}

\appendix

\onecolumn

\section{The Panphasia octree basis functions\label{OCT-DET}}

  In this appendix we define the octree basis functions for \WNF, in
terms of the S$_8$ set of Legendre block functions, which themselves
are defined by eqn~(\ref{legendre_block}).

 Before we can define the octree basis functions themselves we first
define, on the left hand side below, a set of 64 functions each of
which is some linear combination of Legendre block functions
determined by the appropriate matrix equation on the right hand side.
The eight functions, for example $P_{000}$, heading each column on
the left hand side have by construction the functional forms of
the Legendre blocks one level shallower in the octree. The fifty-six
functions below these eight function, will be used below to define
the S$_8$ octree basis functions themselves.
 
%   MACHINE GENERATED

 \begin{equation*}
 \scriptstyle
 \begin{vmatrix}
 \phantom{P_{000}} & \phantom{P_{000}} & \phantom{P_{000}} \\
 &P_{000} &\\
 &Q_{  1} &\\
 &Q_{  2} &\\
 &Q_{  3} &\\
 &Q_{  4} &\\
 &Q_{  5} &\\
 &Q_{  6} &\\
 &Q_{  7} &\\
  & & \\
 \end{vmatrix} \phantom{xx}{\displaystyle =}\phantom{xx}
 \begin{pmatrix}
 \phantom{-c_1} &\phantom{-c_1} &\phantom{-c_1} &\phantom{-c_1} &
 \phantom{-c_1} &\phantom{-c_1} &\phantom{-c_1} &\phantom{-c_1} \\
 \phantom{-}1   & \phantom{-}0   & \phantom{-}0   & \phantom{-}0   & 
 \phantom{-}0   & \phantom{-}0   & \phantom{-}0   & \phantom{-}0     \\
 \phantom{-}0   &           -1   & \phantom{-}0   & \phantom{-}0   & 
 \phantom{-}0   & \phantom{-}0   & \phantom{-}0   & \phantom{-}0     \\
 \phantom{-}0   & \phantom{-}0   &           -1   & \phantom{-}0   & 
 \phantom{-}0   & \phantom{-}0   & \phantom{-}0   & \phantom{-}0     \\
 \phantom{-}0   & \phantom{-}0   & \phantom{-}0   & \phantom{-}1   & 
 \phantom{-}0   & \phantom{-}0   & \phantom{-}0   & \phantom{-}0     \\
 \phantom{-}0   & \phantom{-}0   & \phantom{-}0   & \phantom{-}0   & 
           -1   & \phantom{-}0   & \phantom{-}0   & \phantom{-}0     \\
 \phantom{-}0   & \phantom{-}0   & \phantom{-}0   & \phantom{-}0   & 
 \phantom{-}0   & \phantom{-}1   & \phantom{-}0   & \phantom{-}0     \\
 \phantom{-}0   & \phantom{-}0   & \phantom{-}0   & \phantom{-}0   & 
 \phantom{-}0   & \phantom{-}0   & \phantom{-}1   & \phantom{-}0     \\
 \phantom{-}0   & \phantom{-}0   & \phantom{-}0   & \phantom{-}0   & 
 \phantom{-}0   & \phantom{-}0   & \phantom{-}0   &           -1     \\
  & & & & & & & \\
 \end{pmatrix}
 \phantom{xx}
 \begin{vmatrix}
 \phantom{p_{0000}} & \phantom{p_{0000}} & \phantom{p_{0000}} \\
 &p_{000} &\\
 &p_{001} &\\
 &p_{010} &\\
 &p_{011} &\\
 &p_{100} &\\
 &p_{101} &\\
 &p_{110} &\\
 &p_{111} &\\
  & & \\
 \end{vmatrix}
 \begin{matrix}
 \phantom{xxxxxxxxx}  & \phantom{xxxxxxxxxx}  \\
                             \\
                             \\
                             \\
           Key:              \\
                             \\
                             \\
                             \\
                             \\
  &  \\
 \end{matrix}
 \end{equation*}
 \begin{equation*}
 \scriptstyle
 \begin{vmatrix}
 \phantom{P_{000}} & \phantom{P_{000}} & \phantom{P_{000}} \\
  P_{100}&P_{010}&P_{001}  \\
  Q_{  8}&Q_{ 15}&Q_{ 22}  \\
  Q_{  9}&Q_{ 16}&Q_{ 23}  \\
  Q_{ 11}\text{\footnotemark}&Q_{ 17}&Q_{ 24}  \\
  Q_{ 10}&Q_{ 18}&Q_{ 25}  \\
  Q_{ 12}&Q_{ 19}&Q_{ 26}  \\
  Q_{ 13}&Q_{ 20}&Q_{ 27}  \\
  Q_{ 14}&Q_{ 21}&Q_{ 28}  \\
  & & \\
 \end{vmatrix} \phantom{xx}{\displaystyle =}\phantom{xx}
 \begin{pmatrix}
 \phantom{-c_1} &\phantom{-c_1} &\phantom{-c_1} &\phantom{-c_1} &
 \phantom{-c_1} &\phantom{-c_1} &\phantom{-c_1} &\phantom{-c_1} \\
 \phantom{-}a_{1}&\phantom{-}a_{2}&\phantom{-}0   & \phantom{-}0   & 
 \phantom{-}0   & \phantom{-}0   & \phantom{-}0   & \phantom{-}0     \\
           -a_{2}&\phantom{-}a_{1}&\phantom{-}0   & \phantom{-}0   & 
 \phantom{-}0   & \phantom{-}0   & \phantom{-}0   & \phantom{-}0     \\
 \phantom{-}0   & \phantom{-}0   & \phantom{-}1   & \phantom{-}0   & 
 \phantom{-}0   & \phantom{-}0   & \phantom{-}0   & \phantom{-}0     \\
 \phantom{-}0   & \phantom{-}0   & \phantom{-}0   &           -1   & 
 \phantom{-}0   & \phantom{-}0   & \phantom{-}0   & \phantom{-}0     \\
 \phantom{-}0   & \phantom{-}0   & \phantom{-}0   & \phantom{-}0   & 
 \phantom{-}1   & \phantom{-}0   & \phantom{-}0   & \phantom{-}0     \\
 \phantom{-}0   & \phantom{-}0   & \phantom{-}0   & \phantom{-}0   & 
 \phantom{-}0   &           -1   & \phantom{-}0   & \phantom{-}0     \\
 \phantom{-}0   & \phantom{-}0   & \phantom{-}0   & \phantom{-}0   & 
 \phantom{-}0   & \phantom{-}0   &           -1   & \phantom{-}0     \\
 \phantom{-}0   & \phantom{-}0   & \phantom{-}0   & \phantom{-}0   & 
 \phantom{-}0   & \phantom{-}0   & \phantom{-}0   & \phantom{-}1     \\
  & & & & & & & \\
 \end{pmatrix}
 \phantom{xx}
 \begin{vmatrix}
 \phantom{p_{0000}} & \phantom{p_{0000}} & \phantom{p_{0000}} \\
  p_{000}&p_{000}&p_{000}  \\
  p_{100}&p_{010}&p_{001}  \\
  p_{001}&p_{001}&p_{010}  \\
  p_{011}&p_{011}&p_{011}  \\
  p_{010}&p_{100}&p_{100}  \\
  p_{101}&p_{101}&p_{101}  \\
  p_{110}&p_{110}&p_{110}  \\
  p_{111}&p_{111}&p_{111}  \\
  & & \\
 \end{vmatrix}
 \begin{matrix}
 \phantom{xxxxxxxxx}  & \phantom{xxxxxxxxxx}  \\
                             \\
 a_1 =& \frac{\sqrt{3}}{2}   \\
                             \\
 a_2 =& \frac{1}{2}          \\
                             \\
                             \\
                             \\
                             \\
  &  \\
 \end{matrix}
 \end{equation*}
\footnotetext{It is just a feature of \WNF\ that the ordering of
$Q_{10}$ and $Q_{11}$ is reversed from what might have been expected.}
 \begin{equation*}
 \scriptstyle
 \begin{vmatrix}
 \phantom{P_{000}} & \phantom{P_{000}} & \phantom{P_{000}} \\
  P_{110}&P_{011}&P_{101}  \\
  Q_{ 29}&Q_{ 36}&Q_{ 43}  \\
  Q_{ 30}&Q_{ 37}&Q_{ 44}  \\
  Q_{ 31}&Q_{ 38}&Q_{ 45}  \\
  Q_{ 32}&Q_{ 39}&Q_{ 46}  \\
  Q_{ 33}&Q_{ 40}&Q_{ 47}  \\
  Q_{ 34}&Q_{ 41}&Q_{ 48}  \\
  Q_{ 35}&Q_{ 42}&Q_{ 49}  \\
  & & \\
 \end{vmatrix} \phantom{xx}{\displaystyle =}\phantom{xx}
 \begin{pmatrix}
 \phantom{-c_1} &\phantom{-c_1} &\phantom{-c_1} &\phantom{-c_1} &
 \phantom{-c_1} &\phantom{-c_1} &\phantom{-c_1} &\phantom{-c_1} \\
 \phantom{-}b_{1}&\phantom{-}b_{2}&\phantom{-}b_{2}&\phantom{-}b_{3}&
 \phantom{-}0   & \phantom{-}0   & \phantom{-}0   & \phantom{-}0     \\
           -b_{2}&          -b_{3}&\phantom{-}b_{1}&\phantom{-}b_{2}&
 \phantom{-}0   & \phantom{-}0   & \phantom{-}0   & \phantom{-}0     \\
 \phantom{-}b_{3}&          -b_{2}&\phantom{-}b_{2}&          -b_{1}&
 \phantom{-}0   & \phantom{-}0   & \phantom{-}0   & \phantom{-}0     \\
           -b_{2}&\phantom{-}b_{1}&\phantom{-}b_{3}&          -b_{2}&
 \phantom{-}0   & \phantom{-}0   & \phantom{-}0   & \phantom{-}0     \\
 \phantom{-}0   & \phantom{-}0   & \phantom{-}0   & \phantom{-}0   & 
           -1   & \phantom{-}0   & \phantom{-}0   & \phantom{-}0     \\
 \phantom{-}0   & \phantom{-}0   & \phantom{-}0   & \phantom{-}0   & 
 \phantom{-}0   & \phantom{-}1   & \phantom{-}0   & \phantom{-}0     \\
 \phantom{-}0   & \phantom{-}0   & \phantom{-}0   & \phantom{-}0   & 
 \phantom{-}0   & \phantom{-}0   & \phantom{-}1   & \phantom{-}0     \\
 \phantom{-}0   & \phantom{-}0   & \phantom{-}0   & \phantom{-}0   & 
 \phantom{-}0   & \phantom{-}0   & \phantom{-}0   &           -1     \\
  & & & & & & & \\
 \end{pmatrix}
 \phantom{xx}
 \begin{vmatrix}
 \phantom{p_{0000}} & \phantom{p_{0000}} & \phantom{p_{0000}} \\
  p_{000}&p_{000}&p_{000}  \\
  p_{010}&p_{001}&p_{001}  \\
  p_{100}&p_{010}&p_{100}  \\
  p_{110}&p_{011}&p_{101}  \\
  p_{001}&p_{100}&p_{010}  \\
  p_{011}&p_{101}&p_{011}  \\
  p_{101}&p_{110}&p_{110}  \\
  p_{111}&p_{111}&p_{111}  \\
  & & \\
 \end{vmatrix}
 \begin{matrix}
 \phantom{xxxxxxxxx}  & \phantom{xxxxxxxxxx}  \\
                             \\
 b_1 =& \frac{3}{4}          \\
                             \\
 b_2 =& \frac{\sqrt{3}}{4}   \\
                             \\
 b_3 =& \frac{1}{4}          \\
                             \\
                             \\
  &  \\
 \end{matrix}
 \end{equation*}
 \begin{equation*}
 \scriptstyle
 \begin{vmatrix}
 \phantom{P_{000}} & \phantom{P_{000}} & \phantom{P_{000}} \\
 &P_{111} &\\
 &Q_{ 50} &\\
 &Q_{ 51} &\\
 &Q_{ 52} &\\
 &Q_{ 53} &\\
 &Q_{ 54} &\\
 &Q_{ 55} &\\
 &Q_{ 56} &\\
  & & \\
 \end{vmatrix} \phantom{xx}{\displaystyle =}\phantom{xx}
 \begin{pmatrix}
 \phantom{-c_1} &\phantom{-c_1} &\phantom{-c_1} &\phantom{-c_1} &
 \phantom{-c_1} &\phantom{-c_1} &\phantom{-c_1} &\phantom{-c_1} \\
 \phantom{-}c_{1}&\phantom{-}c_{2}&\phantom{-}c_{2}&\phantom{-}c_{3}&
 \phantom{-}c_{2}&\phantom{-}c_{3}&\phantom{-}c_{3}&\phantom{-}c_{4} \\
           -c_{2}&\phantom{-}c_{1}&\phantom{-}c_{2}&          -c_{3}&
           -c_{2}&\phantom{-}c_{3}&\phantom{-}c_{4}&          -c_{3} \\
           -c_{2}&          -c_{2}&\phantom{-}c_{1}&          -c_{3}&
 \phantom{-}c_{2}&          -c_{4}&\phantom{-}c_{3}&\phantom{-}c_{3} \\
           -c_{2}&\phantom{-}c_{2}&          -c_{2}&\phantom{-}c_{4}&
 \phantom{-}c_{1}&          -c_{3}&\phantom{-}c_{3}&          -c_{3} \\
 \phantom{-}c_{3}&          -c_{3}&          -c_{3}&          -c_{1}&
 \phantom{-}c_{4}&\phantom{-}c_{2}&\phantom{-}c_{2}&          -c_{2} \\
 \phantom{-}c_{3}&\phantom{-}c_{3}&          -c_{4}&          -c_{2}&
           -c_{3}&          -c_{1}&\phantom{-}c_{2}&\phantom{-}c_{2} \\
 \phantom{-}c_{3}&\phantom{-}c_{4}&\phantom{-}c_{3}&          -c_{2}&
 \phantom{-}c_{3}&          -c_{2}&          -c_{1}&          -c_{2} \\
           -c_{4}&\phantom{-}c_{3}&          -c_{3}&          -c_{2}&
 \phantom{-}c_{3}&\phantom{-}c_{2}&          -c_{2}&\phantom{-}c_{1} \\
  & & & & & & & \\
 \end{pmatrix}
 \phantom{xx}
 \begin{vmatrix}
 \phantom{p_{0000}} & \phantom{p_{0000}} & \phantom{p_{0000}} \\
 &p_{000} &\\
 &p_{001} &\\
 &p_{010} &\\
 &p_{011} &\\
 &p_{100} &\\
 &p_{101} &\\
 &p_{110} &\\
 &p_{111} &\\
  & & \\
 \end{vmatrix}
 \begin{matrix}
 \phantom{xxxxxxxxx}  & \phantom{xxxxxxxxxx}  \\
 c_1 =& \sqrt{\frac{27}{64}} \\
                             \\
 c_2 =& \frac{3}{8}          \\
                             \\
 c_3 =& \sqrt{\frac{3}{64}}  \\
                             \\
 c_4 =& \frac{1}{8}          \\
                             \\
  &  \\
 \end{matrix}
 \end{equation*}

%   MACHINE GENERATED 

  The eight Legendre blocks each have distinct symmetries with respect
to a reflection about the thee principal coordinate planes about the
cell centre. So for example $p_{000}$, which is a constant, has even
parity with respect to all three reflections while $p_{111}$ has odd
parity for all three reflections. Before we can define the octree
functions we first define the following function, which is
antisymmetric about the origin:
\begin{equation}
  A(u) = \begin{cases}   \phantom{-}1 &  \text{if $ 0\leq u < 1$;} \\
                                   -1 &  \text{if $ -1 < u < 0$;}  \\
                         \phantom{-}0 &  \text{otherwise.}
         \end{cases}
\end{equation}
 Creating a product of three of these functions each acting on
one of the Cartesian coordinates we arrive at a 
useful function $A^i(x_1)A^j(x_2)A^k(x_3)$, that has the same parities about
the origin as the corresponding Legendre block, $p_{ijk}({\bf x})$,
where ${\bf x} \equiv (x_1,x_2,x_3)$.

 As stated earlier the functions, $P_{ijk}$, heading each of the eight
columns on the left hand side of the matrix equation are linear
combinations of Legendre block functions, and are simply related by
construction to Legendre block functions of twice their linear scale
by the following equation:
\begin{equation}
    p_{ijk}({\bf x}) = P_{ijk}\Big(2|x_1|-\frac{1}{2},
   2|x_2|-\frac{1}{2},2|x_3|-\frac{1}{2}\Big) A^i(x_1)A^j(x_2)A^k(x_3)
\label{lb_def}\end{equation}
similarly taking the remaining seven functions in each of these
columns we can now define the functional forms of the octree basis
functions themselves in analogous way:
\begin{equation}
    q_{n}({\bf x}) = Q_{n}\Big(2|x_1|-\frac{1}{2},
  2|x_2|-\frac{1}{2},2|x_3|-\frac{1}{2}\Big)A^i(x_1)A^j(x_2)A^k(x_3),
\label{octf_def}\end{equation}
  where $n$ is an integer in the inclusive range $[7(i+2j+4k)+1,7(i+2j+4k)+7]$,
and $i$, $j$ and $k$ each take the values of zero or one. 

 Combining all eight columns yields 56 functions, $q_n({\bf x}),
n=1,56$ which gives the functional form for the \WNF\ octree basis
functions, plus the eight Legendre block functions.  These
64 functions when placed in a any given octree cell are mutually orthogonal.
This can be seen as follows. Functions drawn from different columns
have different parities, and are therefore orthogonal.  Within a given
column the orthogonality can be verified by inspection of the
orthogonality between pairs of rows of the square matrices, combined
with the knowledge that the Legendre block functions in the columns on
the right hand side are themselves mutually orthogonal.
 
  Because the 56 octree basis functions are orthogonal not only to
each other, but the 8 Legendre block functions, and the octree basis
functions are built of (smaller) Legendre blocks, then it follows that
any two different octree basis functions, placed in any two
octree cells, are necessarily orthogonal, whether the octree
cells overlap or not.

 Using the fact that the octree basis functions are orthogonal to the
Legendre block functions occupying the same octree cell we can
see this implies that the S$_8$ octree basis functions must have
vanishing zeroth and first moments:
\begin{equation}
 \int  x_1^{\alpha_1} x_2^{\alpha_2} x_3^{\alpha_3}q_n(x_1,x_2,x_3) {\rm d}^3{\bf x} = 0,
\end{equation} for $\alpha_i = 0,1$ where $i=1,2,3$ and the integral is over
all space.

  We can now define the \WNF\ octree basis functions themselves using
the functional forms defined above.  Using the notation established in
Section~\ref{octree}, we define a set of functions for each cell
$(j_1,j_2,j_3)$ in the octree at level $l$:
\begin{equation}
   B^{l,n}_{j_1,j_2,j_3}({\bf x}) = \frac{1}{\Delta_l^{3/2}}q_n
\left(\frac{{\bf x} - {\bf x^c}(l,j_1,j_2,j_3)}{\Delta_l}\right),
\label{OCT_FUN_DEF}\end{equation}
  and $n=1,56$ labels the octree functional forms. 
The terms ${\bf x^c}(l,j_1,_j2,j_3)$ and $\Delta_l$ gives
the cell centre and cell size are defined in eqns~\ref{cell_centre}
and \ref{deltal} respectively. The octree basis
  function obey the following normalisation/orthogonality relations:
\begin{equation}
 \int_{L^3} B^{l_1,n_1}_{j_1,j_2,j_3}({\bf x}) B^{l_2,n_2}_{k_1,k_2,k_3}({\bf x}) {\rm d}^3{\bf x} =
  \delta_{l_1l_2}\delta_{n_1n_2}\delta_{j_1k_1}\delta_{j_2k_2}\delta_{j_3k_3},
\end{equation}
where the integral is over the entire volume of the root cell.
Because they are mutually orthogonal then, as we have shown in
Section~\ref{MATH-INTRO}, the expansion coefficients of a basis
function expansion of a Gaussian white noise field, are independent
Gaussian variables.  In the next Appendix we give the mapping between
the pseudorandom number sequence and the basis functions that defines
the \WNF\ realisation.

\section{Generating a Gaussian pseudorandom sequence and mapping it onto the octree\label{MAP-OCT}}
 As described in Section~\ref{FIND-GEN} we use the MRGk5-93 generator
to provide a very long periodic sequence of pseudorandom numbers $r_i$
which are uniformly distributed between 0 and 1. We use the Box-Muller
transformation \citep{BoxMuller}, with a modification to generate
a corresponding sequence of Gaussian pseudorandom numbers, $g_i$ with
zero mean and unit variance:
\begin{eqnarray}\label{box-muller}
g_{2i\phantom{+1}}   =&    \sqrt{-2\ln(r_{2i})} \cos(2\pi r_{2i+1}) \\
g_{2i+1}             =&    \sqrt{-2\ln(r_{2i})} \sin(2\pi r_{2i+1}). \nonumber
\end{eqnarray}

 A significant fraction of the computing time when evaluating \WNF\
is spent generating the Gaussian pseudorandom numbers. A faster
method is described in \cite{Numerical_recipes92}, but as it uses
the rejection method it is not suitable because we wish to establish
a mapping for the entire available sequence without having to look
at the pseudorandom number values themselves.

 Equation~(\ref{box-muller}) does not work well for the rare occasions
when $r_{2i}$ is very small, as the pseudorandom numbers are discrete
and there is a smallest non-zero pseudorandom number with a magnitude
of about $2.3\times10^{-10}$.  As the number of Gaussian variables in
\WNF\ is huge, the extreme tail of the Gaussian variables will be
truncated if nothing further were done.  To deal with this we first
check if $r_{2i}<10^{-6}$ and if it is replace the value in
eqn~(\ref{box-muller}) with $r_{2i} = 10^{-6}r_{\rm new}$ where
$r_{\rm new}$ is an alternate random number, whose origin will be
discussed below.  Should $r_{\rm new}$ be less than $10^{-6}$ the
procedure is repeated until a larger value is obtained.

 The number $r_{\rm new}$ cannot be taken from the pseudorandom number
sequence close to $r_{2i}$, as with the rejection method, it would
interfere with the mapping between the octree and the pseudorandom
number sequence, or the same number would be reused, which violates
the requirement that the random numbers be independent.  To get round
this $r_{\rm new}$ is computed by advancing the sequence from the
position corresponding to $r_{2i}$ by a very large and arbitrary shift of
$2^{137}+1$. This large shift guarantees in practice that same random
number is not used twice in making some particular initial conditions.
This branching procedure is only really required for the even
pseudorandom numbers,$r_{2i}$, as the odd values $r_{2i+1}$ are used
to calculate an angle which does not have singular behavior at either
end of the range.  However the code used for \WNF\ applies the same
branching conditions to both even and odd values although this makes a
fairly negligible difference to the actual values of the Gaussian pair
for the odd case.

  This whole procedure is only enacted once in a million times.  We
tested that the modified routine does return a Gaussian distribution
well into the tail of the distribution where the modification becomes
important.  The routine was also tested with a less stringent branch
condition, $r_{2i}<10^{-2},$ to ensure that it works if $r_{\rm new}$ is
also small and one or more further iterations is required.

  Having described the origin of the pseudorandom sequence we move onto
the mapping between this sequence and the octree. 
  For each cell in the octree there are 56 octree basis functions and we need to 
generate an expansion coefficient for each of these drawing the value from a
Gaussian distribution with zero mean and unit variance.  For reasons explained
in Section~\ref{OCT-GEN} we also associate a further 8 Gaussian random variables
with each cell. These are not properly part of \WNF\, but can be used if
desired to generate an independent pseudo random field. In total 64 random numbers are needed 
for each octree cell.

 For the ensemble average power spectrum on the scale of the root cell
to be a white noise field we need to add the effects of of an infinite
set of octree basis functions which are larger and overlap the root
cell.  This can be achieved simply by expanding the root cell in
Legendre block functions. The first eight Gaussian pseudorandom
numbers, $g_i, i=0,7$ are reserved for the the coefficients of these
root cell sized Legendre block functions.  After this every 64
consecutive Gaussian pseudorandom numbers are assigned to a particular
octree cell, level by level, with increasing depth. For each cell the
first 56 pseudorandom numbers are assigned to the octree functions,
while the final eight are not part of \WNF\ and are available to
generate a field with one value per octree cell that is independent of
\WNF.  A raster scan pattern over the cells is used at every given level.
Using the same notation for the octree cells as in the previous
appendix we define an integer function:
\begin{equation}
 \phi(l,j_1,j_2,j_3) = 8 + 64\left[4^lj_1 + 2^lj_2 + j_3 + \frac{8^{l-1}-1}{7}\right],
\label{LIN-OCT-ASSIGN}
\end{equation} 
 for $l>0$.   For octree cell  $(j_1,j_2,j_3)$ at level $l$ of the octree the first
and last pseudorandom Gaussian variables for that cell are $g_{[\phi(l,j_1,j_2,j_3)]}$ and
$g_{[\phi(l,j_1,j_2,j_3)+63]}$.

  To define \WNF\ we need also to specify the starting point of the
pseudorandom number sequence.  Table~\ref{list_of_states} gives the
initial state of the random number generator plus some additional
values which are useful cross-checks for anyone wanting to write their
own code to generate \WNF.  The final three entries of the table show
an example where the generation of the two Gaussian variables using
eqn~(\ref{box-muller}) requires the branching procedure described
above to generate the final numbers.

  Section~\ref{PUB-PHASE} described how to publish the phases for a
cuboidal patch within \WNF. The final number in the panphasian
descriptor is a check number.  This check number depends on the
location of the cell in the octree, and also on an ascii character
string included in the descriptor.  For a cuboid at level $l$ of the
octree with a corner nearest the origin at $(j_x,j_y,j_z)$ and
side-lengths $d_x,d_y,d_z$, we define three integers $I_1 =
\phi(l,j_x+d_x-1,j_y,j_z)$, $I_2 = \phi(l,j_x,j_y+d_y-1,j_z)$, $I_3 =
\phi(l,j_x,j_y,j_z+d_z-1)$.

The check number, $N_{\rm check}$ is given by:
\begin{equation}
     N_{\rm check} = \left(T^{I_1}(1) + T^{I_2}(1) + T^{I_3}(1) 
+ \sum_{i=1}^n iT^{{\rm ascii}({\rm string}(i))}(1) \right) \mod m,
\end{equation} 
 where $T$ is the state vector of the random number generator, used in eqn~\ref{def_rand} 
and the sum is over the $n$ characters given in the descriptor name.  The function ascii$()$
returns the ascii value of character. For example: ascii$({\rm A})$ = 65,
 ascii$({\rm a})$ = 97.

\begin{table}
  \begin{center}
{
\begin{tabular}{|l|r|r|r|l|}
\hline
\hline
  $i$          &       $r_i$      &  $g_i$   &  State of pseudorandom number generator for $r_i$
 $\phantom{xxxxxxxxxxxxxxx}$  &  Comment \\
\hline
  0     &     0.716483784  &  0.536408766 & $(1538637210,861452511,1738028090,1398591498, 1039141497)$ & Origin of sequence. \\
  1     &     0.864066010  & -0.615682518 & $(1855567628,1538637210,861452511,1738028090,1398591498)$ &         \\
\hline
 4657948 &  $6.91507\times10^{-8}$ &  4.574061225 &  $(149,1149276986,1622633566,1876117056,1232329462)$ & Branch point \\
 4657949 &   0.101988118         &   3.411353097 &                                        &      \\ 
\hline
4657949+$2^{137}$ &  $8.507921\times10^{-2}$ &      &  $(182706232,  1864678143,  1322192784,   650896850,  1598221492)$ &  $r_{\rm new}$ \\
\hline                                  
\end{tabular}
}
\end{center}
\caption{Some reference values for \WNF.  The first column gives the position of a random number
relative to the starting point.  The sequence $r_i$, are uniformly distributed pseudorandom numbers
in the range $0<r_i<1$.  The sequence $g_i$ are corresponding Gaussian pseudorandom numbers, with
zero mean and unit variance, and are generate in pairs using the Box-Muller transformation as
described in the text of Appendix~\ref{MAP-OCT}  }
\label{list_of_states}
\end{table}

\section{The definition of \WNF.\label{PANPHASIA-DEF}}

  We now combine the results of the previous appendices to give a complete definition
of the \WNF\ field at a position ${\bf x} = (x_1,x_2,x_3)$, where $0 \le x_i < L$, $i=1,2,3$.

\begin{eqnarray*}
\displaystyle
  W_{\rm\WNF}({\bf x}) =  &   \\
  &     \sum\limits_{i_1=0}^1\sum\limits_{i_2=0}^1\sum\limits_{i_3=0}^1 
    {\displaystyle g}_{4i_1+2i_2+i_3}{\displaystyle p}_{i_1i_2i_3}(\frac{2x_1-L}{L},\frac{2x_2-L}{L},\frac{2x_3-L}{L})\\
  &   + \sum\limits_{l=0}^{49} \phantom{x}
        \sum\limits_{j_1=0}^{2^l-1}  \phantom{x}
        \sum\limits_{j_2=0}^{2^l-1}  \phantom{x}
        \sum\limits_{j_3=0}^{2^l-1}  \phantom{x}
        \sum\limits_{n=1}^{56}\phantom{xxx}  {\displaystyle g}_{[\phi(l,j_1,j_2,j_3)+n-1]} B^{l,n}_{j_1,j_2,j_3}({\bf x}),\\
\end{eqnarray*}
 where all the symbols are defined as follows.  The first row on the right hand side is a sum
over the eight Legendre block functions, $p_{j_1j_2j_3}$ defined in eqn~\ref{legendre_block}.  The coefficients
$g_{4i_1+2i_2+i_3}$ are Gaussian pseudorandom numbers with zero mean and unit variance, and are described in
Appendix~\ref{MAP-OCT}.  The summation in the second row on the right hand side is over the entire set of
octree basis functions.  The index $l$ denotes the level in the octree, where $l=1$ corresponds to the root
cell and increases with depth.  The summations over
$j_1,j_2,j_3$ are over the cubic cells making up level $l$ of the octree. A description of the octree is
given in Section~\ref{MATH-INTRO}. The sum over $n$ is over the 56 orthogonal octree basis functions
which occupy each cell and are defined in Appendix~\ref{OCT-DET}.  The sequence of Gaussian
pseudorandom numbers, $g$, are linear within a cell and begin for a given cell at a location given
by the integer function, $\phi(l,j_1,j_2,j_3)$, defined in eqn~\ref{LIN-OCT-ASSIGN}. The octree
basis functions themselves, $ B^{l,n}_{j_1,j_2,j_3}$ are defined by eqn~\ref{OCT_FUN_DEF}. 

\end{document}